\documentclass[a4paper,11pt]{article}

\usepackage{amsmath,amssymb,amsfonts}
\usepackage{calc}
\usepackage{caption2}
\usepackage{hyperref}
\usepackage[pdftex]{color,graphicx}
\usepackage{cite}
\usepackage{multirow}

\numberwithin{equation}{section} 
\setcaptionmargin{1cm}

\long\def\symbolfootnote[#1]#2{\begingroup
\def\thefootnote{\fnsymbol{footnote}}\footnote[#1]{#2}\endgroup}

\graphicspath{{fig/}}
\usepackage{slashed}
\usepackage{amssymb}

\newcommand{\be}{\begin{equation}}
\newcommand{\ee}{\end{equation}}
\newcommand{\bea}{\begin{eqnarray}}
\newcommand{\eea}{\end{eqnarray}}

\usepackage{graphics}
\usepackage{mathrsfs}
\usepackage{amssymb}
\usepackage{verbatim}
\usepackage{float}
\usepackage{cancel}

\def\beq{\begin{equation}}
\def\eeq{\end{equation}}
\def\bea{\begin{eqnarray}}
\def\eea{\end{eqnarray}}
\def\bitem{\begin{itemize}}
\def\eitem{\end{itemize}}

\graphicspath{{Fig/}}

\title{Neutrino physics and $U(1)_R$ lepton number}
\author{Enrico Bertuzzo, Claudia Frugiuele}

\begin{document}
\begin{titlepage}
\vskip 1.0cm
\begin{center}
{\Large \bf  Fitting Neutrino Physics with a $U(1)_R$ Lepton Number  }
\vskip 1.0cm {\large  Enrico Bertuzzo$^{\,a}$, Claudia Frugiuele$^{\,b}$} \\[1cm]
{\it $^{a}$ Institut de Physique Th\'eorique\!
\symbolfootnote[4]{Laboratoire de la Direction des Sciences de la
Mati\`ere du Commissariat \`a l'Energie Atomique et Unit\'e de
Recherche associ\'ee au CNRS (URA 2306).}, CEA-Saclay,\\
  F-91191 Gif-sur-Yvette Cedex, France.}
\\[3mm]
{\it $^{b}$ Ottawa-Carleton Institute for Physics , Department of Physics, Carleton University,\\
1125 Colonel By Drive, Ottawa, Canada, K1S 5B6
}
\vskip 1.0cm
\today
\end{center}

\begin{abstract}
We study neutrino physics in the context of a supersymmetric model where a continuous R-symmetry is identified with the total Lepton Number and one sneutrino can thus 
play the role of the down type Higgs. We show that R-breaking effects communicated to the visible sector by Anomaly Mediation can reproduce neutrino masses and mixing
 solely via radiative contributions, without requiring any additional degree of freedom. In particular, a relatively large reactor angle (as recently observed by the Daya Bay 
collaboration) can be accommodated in ample regions of the parameter space. On the contrary, if the R-breaking is communicated to the visible sector by gravitational effects at the Planck scale, 
additional particles are necessary to accommodate neutrino data.
\end{abstract}
\end{titlepage}
 \section{ Introduction}
 Having already collected an integrated luminosity of $5\;{\rm fb^{-1}}$, the LHC is starting to probe the nature of the (possible) UV completion of the Standard Model 
(SM). Supersymmetry (SUSY) is surely one of the best motivated SM extensions, since it elegantly solves the hierarchy problem. In the construction of the supersymmetric version of the 
SM (SSM), one finds dangerous operators that allow for proton decay. In order to forbid such operators, the common assumption is to enlarge the symmetry group to 
$SU(2)\times U(1)\times G$, where invariance under $G$ forbids proton decay. Typically one assumes that $G$ is a discrete group (R-parity $R_p$) under which ordinary particles are 
even while supersymmetric particles are odd. Beside forbidding Baryon (B) and Lepton (L) number violating operators that generate proton decay (and other flavor changing processes), 
an immediate consequence of R-parity is to make the Lightest 
Supersymmetric Particle (LSP) absolutely stable. Of course there are alternatives to R-parity (\emph{e.g.} one can impose invariance under other discrete groups, under L and/or B, or one can 
extend R-parity to a continuous $U(1)_R$ \cite{Hall:1990dga , Hall:1990hq}), and one can even assume that proton decay is not 
forbidden, as in R-parity violating (RPV) theories (see \cite{Barbier:2004ez} for a comprehensive review), where however the coefficients of the L and B violating operators must be strongly 
suppressed.
\par
The case of $G=U(1)_R$, the continuous group that contains R-parity as $Z_2$ subgroup, requires to go beyond the minimal scenario. Indeed, the R-symmetry forbids Majorana gaugino masses, 
but Dirac gaugino masses are 
allowed if the gauge sector of the theory is enlarged to the one of $N=2$ SUSY, \emph{e.g.} including adjoint superfields $\Phi^a_i$ for each gauge group $G_i$. If SUSY breaking 
is transmitted to the visible sector through a spurion D-term, $\langle W'_\alpha\rangle = D' \theta_\alpha$, then a 
lagrangian term of the form $\frac{1}{M}\int d^2 \theta (W' W_i^a) \Phi^a_i$ generates Dirac mass terms of order $m_d \sim \frac{D'}{M}$. 
 R-symmetric models \cite{Kribs:2007ac, Davies:2011mp, Frugiuele:2011mh} represent an interesting possibility to explore for several reasons.
First of all, gaugino one loop contributions to squared soft masses are finite \cite{Fox:2002bu}, so that the fine-tuning issue
for the gluino is softened (see \emph{e.g.} \cite{Papucci:2011wy}). In addition, the LHC phenomenology is non standard, both due to the Dirac nature of the 
gluino \cite{Benakli:2011vb, Freitas:2009dp, Heikinheimo:2011fk, Kribs:2012gx} and to the presence of additional particles that can be rather easily detected \cite{Choi:2009jc}. 
Moreover, the Flavor Problem is also softened, since unsuppressed flavor changing terms are now allowed for sufficiently heavy gaugino Dirac masses \cite{Kribs:2007ac, Fok:2010vk}.
\par
As it has been recently explored in \cite{Frugiuele:2011mh} and \cite{Brust:2011tb} \footnote{See also \cite{fayet}  and \cite{Gherghetta:2003wm} for earlier attempts.}, it is not necessary to define the R-symmetry as the 
continuous symmetry containing R-parity. Indeed,
proton stability could be ensured  also identifying the R-symmetry with Lepton number \cite{Frugiuele:2011mh} or with Baryon number \cite{Brust:2011tb}.
Both scenarios violate $R_p$, but proton stability is guaranteed without any suppressed coupling, since the model posses either an accidental standard Baryon or standard Lepton number.
We will focus here on a scenario where the R-symmetry is identified with Lepton number.
One of the distinctive feature of this idea is that it allows for a sneutrino to play the role of down-type Higgs \cite{Frugiuele:2011mh}\footnote{The idea of having a non zero sneutrino vev has been 
extensively explored in the literature, see \cite{sneutrinovev} for examples.}.
Assuming the R-symmetry not to be spontaneously broken by the sneutrino vev, one is then forced to require vanishing Lepton number for the slepton doublet. This immediately implies that 
neutrino Majorana masses are forbidden by the R-symmetry and that the sneutrino vev, being unrelated to neutrino masses, can be large enough to give mass to the bottom quark.
\par
However, the R symmetry is not an exact symmetry, since an irreducible source of R- breaking ($\cancel{R}$ from now on) is given by the gravitino mass necessary to cancel the 
cosmological constant. This suggests a tight connection between neutrino physics and SUSY breaking. 
\par
In the specific model presented in \cite{Frugiuele:2011mh}, the R-symmetry was identified with the lepton number of a specific flavor, and only one non zero neutrino mass was generated. 
We want here to enlarge the R-symmetry to the total Lepton number to see whether this more realistic scenario can reproduce neutrino physics, analyzing in detail the parameter space 
compatible with the present experimental neutrino data.

\section{$R$-symmetry as  global lepton number}\label{Sec:R-symm_as_Lepton_numb}
 \begin{table}\label{Table:particle_content}
\centering 
{
\begin{tabular}{l|l|l} 
\multicolumn{1}{c|}{\textbf{SuperField}} & 
\multicolumn{1}{c|}{\textbf{$ ( SU(3)_c,SU(2)_L)_{U(1)_Y} $}} & 
\multicolumn{1}{c}{\textbf{$U(1)_R$}} \\ 
\hline 
$Q$ & \  \ \ \ \ \ \ $ (3,2)_{\frac{1}{6}}$ \ \ & 1 \\ 
$U^c_i$ & \  \ \ \ \ \ \ $ (\bar 3,1)_{-\frac{2}{3}}$ \ \ & 1 \\  
$D^c_i$ & \  \ \ \ \ \ \ $ (\bar 3,1)_{\frac{1}{3}}$ \ \ & 1 \\ 
$E^c$ & \  \ \ \ \ \ \ $ (1,1)_{1}$ \ \ & 2 \\ 
$L$ & \  \ \ \ \ \ \ $ (1,2)_{-\frac{1}{2}}$ \ \ & 0 \\ 
$H_u$ & \  \ \ \ \ \ \ $ (1,2)_{\frac{1}{2}}$ \ \ & 0 \\
$R_d$ & \  \ \ \ \ \ \ $ (1,2)_{-\frac{1}{2}}$ \ \ & 2 \\ 
$ \Phi_{\tilde W} $ &  \  \ \ \ \ \ \ $ (1,3)_0$ & 0 \\ 
$ \Phi_{\tilde B} $ &  \  \ \ \ \ \ \ $ (1,1)_0$ & 0 \\ 
$ \Phi_{\tilde g}$ &  \  \ \ \ \ \ \ $ (8,1)_0$ & 0\\
\end{tabular} 
}\qquad\qquad 
\caption{R-charge assignment for the chiral supermultiplets in our model.} 
\label{table:particlecontent}
\end{table} 
Let us now describe our framework. We generalize the model of \cite{Frugiuele:2011mh} in such a way that the R-symmetry is identified with the global lepton number, 
$\mathrm{U(1)}_R=U(1)_L$. In particular, all the R-charges of Lepton doublets and singlets are respectively fixed to $0$ and $2$, see Table \ref{Table:particle_content}. 
The $R_d$ electroweak doublet with R-charge $2$, introduced to have an anomaly free framework, will play the role of an inert doublet (since we do not want the R-symmetry to be spontaneously broken), 
while the role of the usual down-type Higgs doublet 
will be played by a combination of sleptons, as we will explain later on. Since R-symmetry invariance is incompatible with Majorana gaugino masses, it is necessary to introduce 
three adjoint superfields, 
$\Phi_{\tilde W, \tilde B, \tilde g}$, that couple to the ordinary gauginos via D-term SUSY breaking \cite{Fox:2002bu} to generate Dirac masses.
\par
The most general superpotential compatible with the given R-charge assignment is:
\bea\label{eq:W-1}
W &=& \mu H_u R_d+ H_u Q Y_U U^c +\sum_{ijk}{ \lambda_{ijk} L_i L_j E^c_k}+ \nonumber\\
 && + \sum_{ijk}{ \lambda'_{ijk} L_i Q_j D^c_k}+ \lambda_S H_u \Phi_{\tilde B} R_d+\lambda_T H_u \Phi_{\tilde W} R_d.
\eea
where  $\lambda_{ijk}=-\lambda_{jik}$ from the antisymmetry of $L_iL_j$.\\
The  R-conserving SUSY breaking soft lagrangian is instead:
\bea\label{eq:LSSB_R_cons}
{\cal L}_R & =&  m^2_{H_u} h^\dagger_u h_u+ \sum_{ij}{ (m^2_{Lij} \tilde \ell_i^{\dagger} \tilde \ell_j+m^2_{Rij} \tilde e_i^{\dagger} \tilde e_j)}- \sum_{i}{  b_{\mu}^i h_u \tilde \ell_i}+ \nonumber \\  
& &+ \sum_{ij}{ (m^2_{qij} \tilde q_i^{\dagger} \tilde q_j+m^2_{dij} \tilde d_i^{\dagger} \tilde d_j+m^2_{uij} \tilde u_i^{\dagger} \tilde u_j)} + M_B \tilde{B} \tilde{\psi}_{\tilde B} + 
M_{\tilde W} {\rm tr}(\tilde{W} \tilde{\psi}_{\tilde W})\;.
\eea
The R-symmetry cannot be an exact symmetry, since it is broken at least by the gravitino mass necessary to cancel the cosmological constant. 
To write down the $\cancel{R}$ soft SUSY breaking lagrangian, we need an ansatz on how the R-breaking is communicated to the visible sector. A minimal scenario is to assume that gravity 
conserves the R-symmetry \cite{Kribs:2010md}, so that R-breaking effects are communicated to the visible sector only through Anomaly Mediation; however, we can also imagine that gravity effects 
at the Planck scale can break the R-symmetry.
\par
In the first case, which we will call Anomaly Mediation R-Breaking (AMRB) scenario, the soft R-breaking lagrangian is given by:
\be\label{eq:AMRB_L}
\mathcal{L}^{{\rm AMRB}}_{\cancel{R}} =  \mathcal{L}_{{\rm Majorana}}+ \mathcal{L}_{{\rm A}}+ B_{\mu} h_u r_d
\ee
where
\bea\label{eq:softAMRB}
\mathcal{L}_{{\rm Majorana}} &=&  m_{B} \tilde B \tilde B + m_{\tilde W} {\rm tr}(\tilde W \tilde W) +  m_{g} {\rm tr}(\tilde g \tilde g) \; , \nonumber\\
{\cal L}_{A} &=& A^\lambda_{ijk} \tilde{\ell}_i \tilde{\ell}_j \tilde{e}^c_k + A^D_{ijk} \tilde{\ell}_i \tilde{q}_j \tilde{d}^c_k + h_u \tilde{q} A^U \tilde{u}^c.
\eea
The first term contains gaugino Majorana masses of order $m\simeq \frac{m_{3/2}}{16 \pi^2}$ \footnote{See \cite{Gherghetta:1999sw} for the exact expressions of gaugino Majorana masses and A-terms in 
Anomaly Mediation.}, while the second one contains trilinear scalar interactions proportional to the supersymmetric Yukawa couplings.
\par
Turning to the case in which gravitational effects at the Planck scale break the R-symmetry (which we will call Planck Mediated R-Breaking (PMRB) scenario), the R-breaking structure is much 
richer than in the previous case, since now all the operators suppressed by some power of the Planck scale can contribute. 
The R-conserving superpotential and soft SUSY breaking lagrangian, Eqs.(\ref{eq:W-1}), (\ref{eq:LSSB_R_cons}), are corrected by the following $\cancel{R}$-contributions:
\bea\label{eq:PBRM}
W^{\rm PMRB}_{\cancel{R}}  &=&  \sum_i\mu_i H_u L_i  +  \frac{1}{2} m_T {\rm tr}(\Phi_{\tilde W} \Phi_{\tilde W})+ \frac{1}{2} m_S \Phi_{\tilde B} \Phi_{\tilde B}\; , \nonumber\\
\mathcal{L}^{{\rm PMRB}}_{\cancel{R}} &=& \mathcal{L}_{{\rm Majorana}}+ \mathcal{L}_{{\rm A}}+ B_{\mu} h_u r_d.
\eea
The $\cancel{R}$ soft SUSY breaking contribution has the same structure as in Eq. (\ref{eq:softAMRB}), but now we simply expect all the terms generated to be of order 
of the gravitino mass $m_{3/2}$ and the A-terms not to be aligned to the supersymmetric Yukawa couplings. Let us notice the appearance of $\mu$-terms and Majorana masses for the Adjoint Fermions, also of order $m_{3/2}$. 
As we will see, they will play an essential role in neutrino physics.
\par
Let us now study how electroweak symmetry breaking works in this framework and how fermions get masses.
Since all the sleptons have a $b_{\mu} $ term, Eq. (\ref{eq:LSSB_R_cons}), in a general basis all sneutrinos will get  a vev.
However, we can use the freedom to rotate slepton fields to work in a ``single vev basis''  where just one sneutrino gets a vev \footnote{This is similar to what happens in RPV SUSY 
models, \cite{Barbier:2004ez}.}. 
We will denote with ${A,B,C} $ the flavor indexes in this basis, with $\tilde{L}_A$ referring to the doublet that plays the role of the down-type Higgs.
The  superpotential can be rewritten as:
\be\label{eq:W_R_cons_single_vev}
W=  \mu H_u R_d+ \lambda_S H_u \Phi_{\tilde B} R_d+\lambda_T H_u \Phi_{\tilde{W}} R_d+ W_{{\rm Yukawa}} + W_{{\rm trilinear}} ,
\ee
with
\bea\label{eq:single_vev_Yuk_tril}
 W_{{\rm Yukawa}} &=& y_B L_A L_B E^c_B+ y_C L_A L_C E^c_C+ y^D_i L_A Q_i D_i^c+ H_u Q Y^U U^c,\nonumber\\
 W_{{\rm trilinear}} &=& \sum_{i=A,B,C}\lambda_{BCi} L_B L_C E^c_i+ \sum_{ij}{( \lambda'_{Bij} L_B Q_i D_j^c+ \lambda'_{Cij} L_C Q_i D_j^c)},
\eea
where $y_B \equiv \lambda_{ABB}$ and $y_C \equiv \lambda_{ACC}$. In the new basis, the R-conserving soft lagrangian of Eq. (\ref{eq:LSSB_R_cons}) maintains the same form, while the 
R-breaking ones of Eqs. 
(\ref{eq:softAMRB}),(\ref{eq:PBRM}) now read
\be\label{eq:AMRB_L2}
{\cal L}_{\rm A} =  \mathcal{L}_{{\rm LR}}+ \mathcal{L}_{{\rm trilinear}},
\ee
with
\bea\label{eq:softAMRB2}
\mathcal{L}_{{\rm LR}} &=&  A_B \tilde \ell_A  \tilde \ell_B  \tilde e^c_B+ A_C \tilde \ell_A  \tilde \ell_C \tilde e^c_C +A_{BC} \tilde \ell_A  \tilde \ell_B \tilde e^c_C+A_{CB} \tilde \ell_A  \tilde \ell_C \tilde e^c_B A^D_i \tilde \ell_A  \tilde q_i  \tilde  d^c_i+  
h_u  \tilde q A^U  \tilde u^c_i, \nonumber\\
\mathcal{L}_{\rm trilinear} &=&  A_{BCi} \tilde \ell_B  \tilde \ell_C  \tilde e^c_i + A'_{Bij} \tilde \ell_B  \tilde q_i  \tilde d^c_j + A'_{Cij} \tilde \ell_C  \tilde q_i  \tilde d^c_j.
\eea
The first term gives slepton and squark left-right mixing \footnote{In the AMRB scenario the off diagonal terms $A_{BC}$, $A_{CB}$ are zero.}, while the second term contains trilinear scalar interactions that do not involve the slepton that 
takes vev. Let us stress that the gaugino Majorana masses and the scalar left/right mixing will play a crucial role in the generation of neutrino masses.
\par
The analysis of the scalar potential can be done along the line of Ref. \cite{Barbier:2004ez}, although in our case the situation is more involved. Indeed, when the left handed slepton 
soft squared mass matrix is not flavor universal, a mixing between the sneutrino that takes vev and the other two is in principle possible, so that we expect the physical Higgs to be an 
admixture of all 
the three sneutrinos. On the contrary, when the squared mass matrix is flavor universal, the resulting scalar potential is the usual one \cite{Frugiuele:2011mh}. We assume 
here for simplicity that, at leading order, the soft squared mass matrix is flavor universal, deferring to a future work 
the analysis of the non flavor universal case.
\par
From Eq. (\ref{eq:single_vev_Yuk_tril}) it is immediate to notice that the charged lepton
of flavor $A$ cannot acquire mass trough  a SUSY invariant Yukawa term as the operator $ \ell_A \ell_A e^c_A$ is null due to the $SU(2)$ invariance.
Therefore, a mass for the lepton $\ell_A $ must be generated by a hard SUSY breaking sector through couplings between messengers and leptonic superfields \cite{Frugiuele:2011mh}.
However,  in the present scenario, this sector will generate hard Yukawa couplings also for the $B$ and $C$ flavors.\\
If we assume that the main contribution to $\ell_{B,C}$ masses comes from the supersymmetric Yukawa couplings, the additional contribution from the hard sector 
must somehow be suppressed. This makes  $A=e$ the simplest possibility.
Indeed, if $A=\tau$, the $\tau$ lepton  mass must be generated by the hard sector, while the hard contribution to the other masses must be suppressed 
(for example requiring the hard Yukawa couplings $y_{ij}$ to satisfy $y_{ij} \ll 10^{-6}$). 
This corresponds to assuming a large hierarchy between the hard Yukawa couplings. The same line of reasoning can be applied in the $A=\mu$ case. 
If instead $A=e$, a hard Yukawa contribution which generates Yukawa couplings of order $y_e \simeq {\cal O}(10^{-6})$ for all the charged leptons does not give a too large contribution to the 
$\mu$ and $\tau$ masses, while providing the correct order of magnitude for an electron mass. Since in this case there is no need to introduce any large hierarchy in the new sector, 
it appears a more natural choice. 
A possible example of hard Yukawa sector is given in \cite{Frugiuele:2011mh}; however, let us stress that, since we will left largely undetermined this sector, in what follows we will 
analyze also the 
cases in which $A\neq e$.
\par
As a  last comment, let us stress that the interaction terms of $W_{{\rm trilinear}}$ (which are not present in \cite{Frugiuele:2011mh}) closely resemble the trilinear interaction terms that appear in RPV theories \cite{Barbier:2004ez}. However, in 
our case all the off diagonal terms involving the flavor $A$, $\lambda^{(')}_{Aij}$, are zero in the single vev basis, so that the number of parameters is reduced. 
Moreover, coupling of the type $\lambda^{(')}_{Aii}$ now play the role of Yukawa couplings and are not free parameters. We conclude that our scenario is a variation of RPV models (with 
less parameters), although as we will see a larger amount of R-parity violation in the neutrino sector than in the standard case will be allowed.

\subsection{Electroweak precision measurements  and flavor constraints}\label{sec:exp-constr}

Let us now discuss in turn the experimental constraints coming from Electroweak Precision Measurements (EWPM) and from flavor physics.
\par
One of the distinctive feature of models where the R-symmetry is identified with Lepton Number, is that all the supersymmetric partners, with the exception of charged sleptons and sneutrinos, have a 
non vanishing lepton number. 
As a consequence, charged leptons and neutrinos can mix with the ``new'' spin $1/2$ leptons (Dirac gauginos and higgsinos).
A priori, the neutralino mass matrix is a $9\times 9$ squared matrix, while the chargino mass matrix is a $12\times 12$ square matrix. 
However, in the single vev basis, the leptons of flavors $B$ and $C$ do not mix with any other fermion, so that the effective matrix is the same as in \cite{Frugiuele:2011mh}, to which we refer 
for a detailed analysis of the mass eigenstates. The important point to stress for our purpose is  that in the R-symmetric limit  all neutrinos are massless. 
Also, the same bounds on the sneutrino vev 
coming from the bounds on the coupling of the $Z$ boson to charged leptons apply, \emph{i.e.} for $ M_{\tilde W} \sim 1\;{\rm TeV}$ one should have $v_A \lesssim 40\;{\rm GeV}$.
\par
Let us now turn to the bounds on trilinear couplings appearing in $W_{\rm Yukawa}$ and $W_{\rm trilinear}$ \cite{Dreiner:2006gu}. Since these are RPV couplings, we refer to 
\cite{Barbier:2004ez, Dreiner:2006gu} for a detailed 
description of the origin of the various bounds. It is interesting to notice that our framework has distinctive differences both with the model of \cite{Frugiuele:2011mh} and with the standard RPV SUSY.
\par
On the one hand, Lepton Flavor Violating (LFV) processes are allowed in our framework but not in \cite{Frugiuele:2011mh}, and the same is true also for semileptonic meson decays 
(such as rare decays of $B$ and $ K$ mesons), unless we assume alignment between the matrices $(\lambda'_{B,C})_{ij}$ and the quark mass matrix. 
\par
On the other hand, even though our situation 
is more similar to the standard RPV SUSY, some bounds have a different interpretation. In particular, bounds that involve a product between two trilinear couplings can now 
involve one Yukawa coupling. In order to maximise the parameter space for the sneutrino vev we read these bounds as vev dependent constraints on the trilinear couplings appearing
in $W_{\rm trilinear}$.
 For example, when $A=e $, the LFV process $\mu \rightarrow e \gamma$ 
puts a bound $|\lambda^{*}_{233} \lambda_{133}| \lesssim  2.3 \times 10^{-4} \left(\frac{m_{\tilde{\ell}_L}}{100\; {\rm GeV}}\right)^2$ (assuming for simplicity degenerate slepton masses).   
Since in our model $\lambda_{133} = \frac{m_\tau}{v_e}$ is the $\tau$ Yukawa coupling, the bound can be restated as 
$|\lambda_{233}| \lesssim  2.3 \times 10^{-4} \frac{ v_e}{ m_{\tau}} \left(\frac{m_{\tilde{\ell}_L}}{100\; {\rm GeV}}\right)^2 \simeq 0.002-0.07$  for  
$m_{ \tilde{\ell}_{L}} \simeq 200\;{\rm GeV}$ and $v_e=(10-80)\;{\rm GeV}$. At the same time, it is true that, among the constraints that involve only one trilinear coupling and not 
a product, some will refer to bounds on Yukawa couplings, implying thus a bound on the sneutrino vev.
\par
In the following we will always assume that trilinear couplings not directly related to neutrino physics are always small enough to satisfy all the experimental constraints.

\section{Neutrino physics and $U(1)_R $ lepton number}

In  our model the R-symmetry is identified with the global Lepton number, so that $U(1)_R$ breaking corresponds to Lepton Number breaking.
In the following section we will discuss how neutrino masses and mixing are generated from R-symmetry breaking effects. 
The problem of neutrino masses in models with an R-symmetry have been 
studied both for Majorana \cite{Kumar:2009sf} and Dirac \cite{Davies:2011js} neutrinos. 
Both scenarios require to enlarge the particle content of the model introducing right handed neutrinos.
 Indeed, in the standard R-symmetric scenario  \cite{Kribs:2007ac, Davies:2011mp}, there is no natural connection between the R-breaking and Majorana 
neutrino masses, since these are allowed by R-symmetry (all lepton superfields have R-charge $1$). A priori, however, R-symmetry does not forbid Dirac masses either, since their presence 
depend on the R-charge assignment of right-handed neutrinos. This makes the connection between neutrino Dirac masses and R-symmetry breaking less stringent.
\par
On the contrary, in our scenario there is a clear connection between Majorana neutrino masses and R-breaking effects, since such Majorana masses are clearly incompatible with the $U(1)_R$ 
symmetry. In this way, in principle we don't need to introduce any additional particle (\emph{i.e.} right-handed neutrinos) in order to generate non zero masses. While, as we 
will see, this will be true 
for AMRB, in the case of PMRB additional structure will be necessary in order to reproduce neutrino masses and mixing, making this scenario less compelling.
\par
Let us stress again that in our scenario the scale at which Lepton Number is broken is deeply connected with the scale of supersymmetry breaking through the gravitino mass, 
while in general the Majorana neutrino masses generated through the Weinberg operator call for a very large scale, which may or may not be connected to the scale of supersymmetry 
breaking.

\subsection{Neutrino masses and mixings}

\begin{table}[tb]
\begin{center}
\begin{tabular}{|c|c|c|}
  \hline
  Quantity & ref. \cite{Fogli:2011qn} & ref. \cite{Schwetz:2011qt}\cite{Schwetz:2011zk}\\
  \hline
  $\Delta m^2_{sun}~(10^{-5}~{\rm eV}^2)$                  & $7.58^{+0.22}_{-0.26}$                  & $7.59^{+0.20}_{-0.18}$  \\
  \hline
  \multirow{2}{*}{$\Delta m^2_{atm}~(10^{-3}~{\rm eV}^2)$} & \multirow{2}{*}{$2.35^{+0.12}_{-0.09}$} & $2.50^{+0.09}_{-0.16}$  \\
                                                           &                                         & $-(2.40^{+0.08}_{-0.09})$  \\
  \hline
  $\sin^2\theta_{12}$                                      &      $0.312^{+0.017}_{-0.016}$          & $0.312^{+0.017}_{-0.015}$ \\
  \hline
  $\sin^2\theta_{23}$                                      & $0.42^{+0.08}_{-0.03}$                  &  $0.52^{+0.06}_{-0.06}$ \\
  \hline
  \multirow{2}{*}{$\sin^2\theta_{13}$}                     & \multirow{2}{*}{$0.025\pm0.007$}        &$0.013^{+0.007}_{-0.005}$  \\
                                                           &                                         &$0.016^{+0.008}_{-0.006}$  \\
  \hline
  \end{tabular}
\end{center}
\caption{\label{tab:neutrino_data} Fits to neutrino oscillation data. Where two different values are present for one parameter, upper and lower row refer respectively to Normal and 
Inverted Hierarchy.}
\end{table}

Before analyzing the neutrino phenomenology in our framework, let us briefly summarize some features of a general neutrino mass matrix.
\par
As it is well known, the neutrino mass matrix is largely undetermined, since we lack of information on the absolute neutrino mass scale and on the hierarchy between the mass 
eigenstates. For three active neutrinos, the present data are summarized in Table \ref{tab:neutrino_data}.
\par
At the same time, CMB data point towards $\sum_i m_{\nu,i} \lesssim 0.6\;{\rm eV}$ (see \emph{e.g.} \cite{Abazajian:2011dt}), from which one can infer a loose upper bound 
$m_{lightest} \lesssim 0.1\;{\rm eV}$ for both hierarchies. 
Using data in the expression of the neutrino mass matrix in terms of masses and mixing, we expect the following general form for the mass matrix (in the $(\nu_e, \nu_\mu, \nu_\tau)$ basis):
\begin{description}
 \item[(*)] Normal Hierarchy:
\be\label{eq:phen_NH}
M_\nu \simeq  m_\nu^{small}
\begin{pmatrix}
{\cal O}(\varepsilon) & {\cal O}(\varepsilon)   & {\cal O}(\varepsilon^2) \cr
{\cal O}(\varepsilon) & {\cal O}(1)             & {\cal O}(1)           \cr
{\cal O}(\varepsilon) & {\cal O}(1)             & {\cal O}(1)
\end{pmatrix}\;,~~~
M_\nu \simeq  m_\nu^{large }
\begin{pmatrix}
{\cal O}(1)           & {\cal O}(\epsilon)      & {\cal O}(\epsilon)       \cr
{\cal O}(\epsilon)    & {\cal O}(1)             & {\cal O}(\sqrt\epsilon)  \cr
{\cal O}(\epsilon)    & {\cal O}(\sqrt\epsilon) & {\cal O}(1)
\end{pmatrix}\;;
\ee
 \item[(*)] Inverted Hierarchy:
\be\label{eq:phen_IH}
M_\nu \simeq  m_\nu^{small}
\begin{pmatrix}
{\cal O}(1)           & {\cal O}(\varepsilon)   & {\cal O}(\varepsilon) \cr
{\cal O}(\varepsilon) & {\cal O}(1)             & {\cal O}(1)           \cr
{\cal O}(\varepsilon) & {\cal O}(1)             & {\cal O}(1)
\end{pmatrix},~~~
M_\nu \simeq  m_\nu^{large }
\begin{pmatrix}
{\cal O}(1)           & {\cal O}(\epsilon)      & {\cal O}(\epsilon)       \cr
{\cal O}(\epsilon)    & {\cal O}(1)             & {\cal O}(\sqrt\epsilon)  \cr
{\cal O}(\epsilon)    & {\cal O}(\sqrt\epsilon) & {\cal O}(1)
\end{pmatrix};
\ee
\end{description}
The matrices on the left and on the right refer respectively to a small (of order ${\cal{O}}(10^{-5}\;\mathrm{eV})$) and large (of order ${\cal{O}}(10^{-1}\;\mathrm{eV})$) 
lightest neutrino mass. 
The exact value of the coefficients depends on the chosen values of neutrino masses and mixing angles (we discard here the dependence on Dirac and Majorana phases); however, 
typical order of magnitudes for the elements are
$$
m_\nu^{small} \simeq {\cal{O}}(10^{-2}\;\mathrm{eV}),~~ m_\nu^{large} \simeq {\cal{O}}(10^{-1}\;\mathrm{eV}),~~\varepsilon \simeq {\cal{O}}(10^{-1}\;\mathrm{eV}),~~
\epsilon \simeq {\cal{O}}(10^{-3}\;\mathrm{eV})
$$
In what follows, taking the approach of \cite{Bhattacharyya:2011zv}, we will focus on specific forms for the neutrino mass matrix that we consider representative of the different 
phenomenological scenarios. In particular,  we will focus on the two following matrices, representative respectively of the Normal and Inverted Hierarchy cases for small lightest 
neutrino mass:
\be\label{eq:exp_matr_NH}
 m_\nu^{\mathrm{NH}} \simeq 10^{-2}
\begin{pmatrix}
0.39   & 0.80   & 0.017   \cr
0.80   & 2.99   & 2.08    \cr
0.017  & 2.08   & 2.49
\end{pmatrix}\;\mathrm{eV}\; ;
~~~ m_\nu^{\mathrm{IH}} \simeq 10^{-2}
\begin{pmatrix}
 4.7 	& -0.54 & -0.52    \cr
-0.54 	& 2.19 	& -2.36	   \cr
-0.52 	& -2.36 & 2.8
\end{pmatrix}\;\mathrm{eV}\;.
\ee
We do not show here the corresponding matrices for the large lightest neutrino mass scenario because, as we will explain later on, they can be reproduced only in a very small region of parameter 
space. 
To construct the previous matrices, we fixed the lightest neutrino mass to $2\times 10^{-5}\;\mathrm{eV}$, while the other parameters are fixed as follows: 
$\Delta m^2_{12}\simeq 7.6\times 10^{-5}\;\mathrm{eV}^2$, 
$\Delta m^2_{13}\simeq 2.4\times 10^{-3}\;\mathrm{eV}^2$, $\sin^2\theta_{12}\simeq 0.3$, $\sin^2\theta_{23}\simeq0.47$, $\sin^2\theta_{13}\simeq0.024$, \emph{i.e.} we take 
$\theta_{13} \simeq 9^\circ$ as recently observed by the Daya Bay collaboration \cite{An:2012eh}. For simplicity, we have also assumed a vanishing CP violating phase.

\begin{center}
 \begin{figure}[tb]
  \begin{tabular}{cc}
   \includegraphics[width=0.45\textwidth]{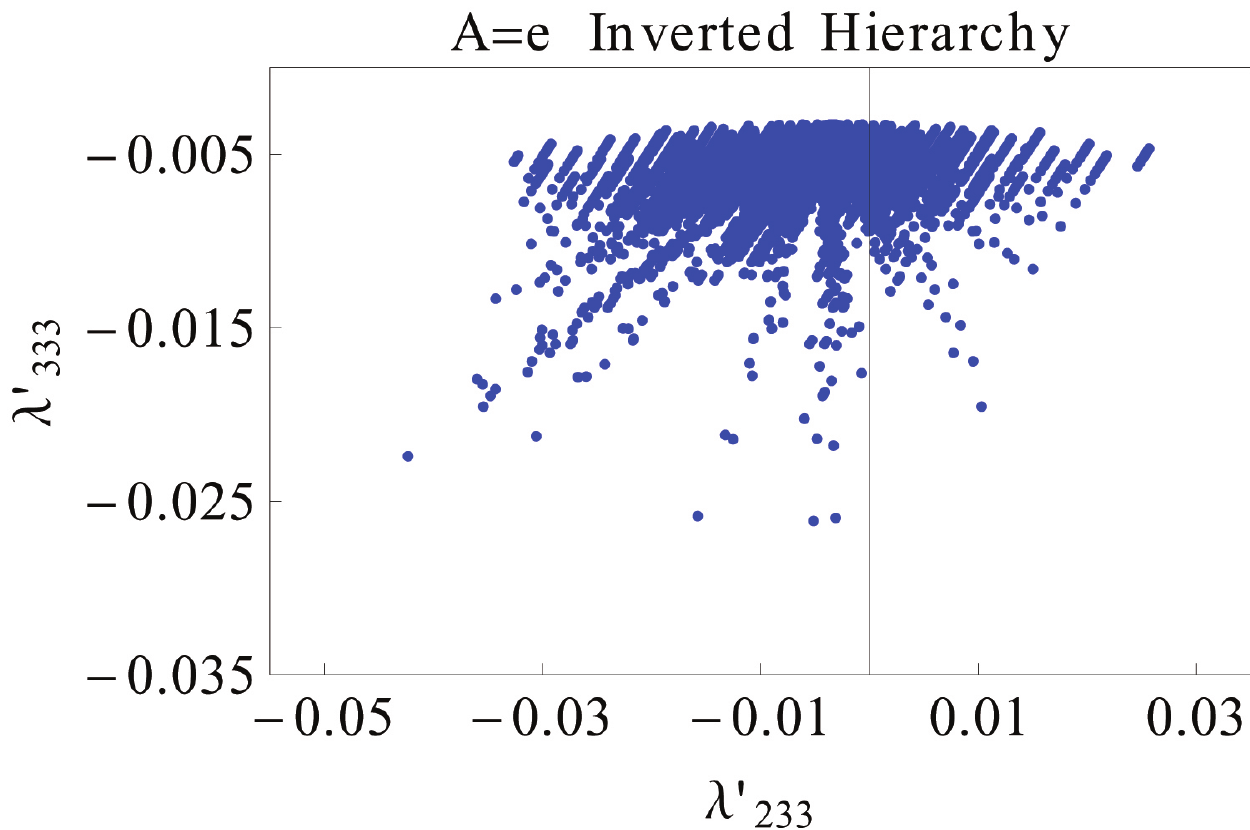}         & \includegraphics[width=0.42\textwidth]{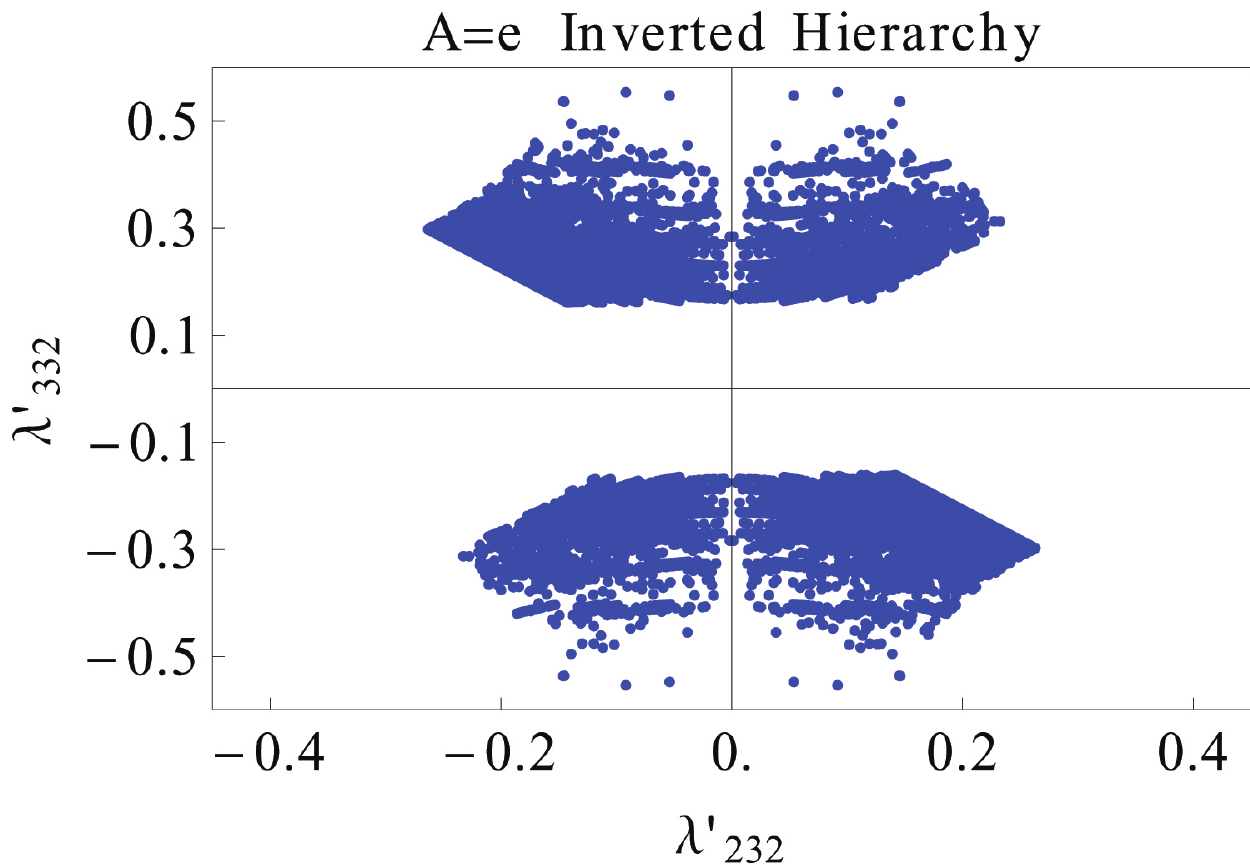} \\
   \includegraphics[width=0.45\textwidth]{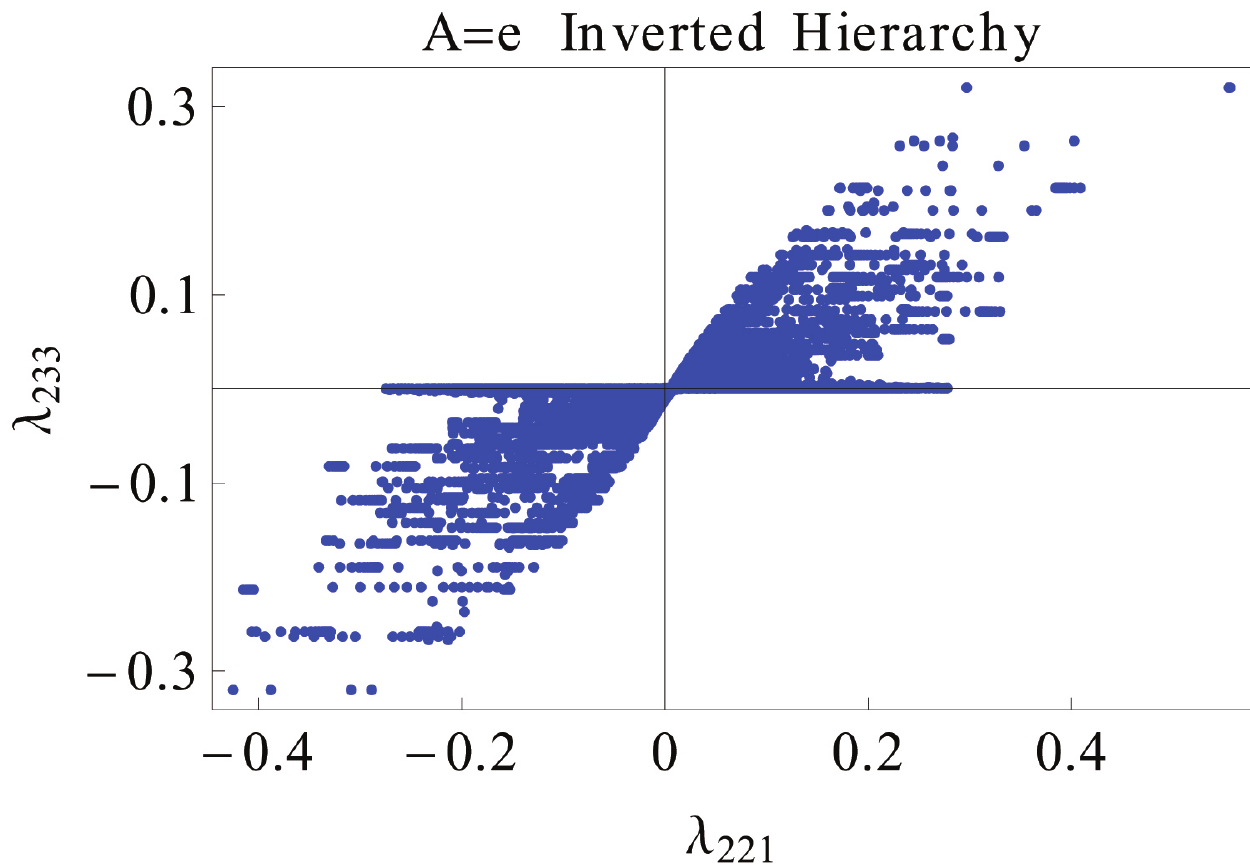}         & \includegraphics[width=0.42\textwidth]{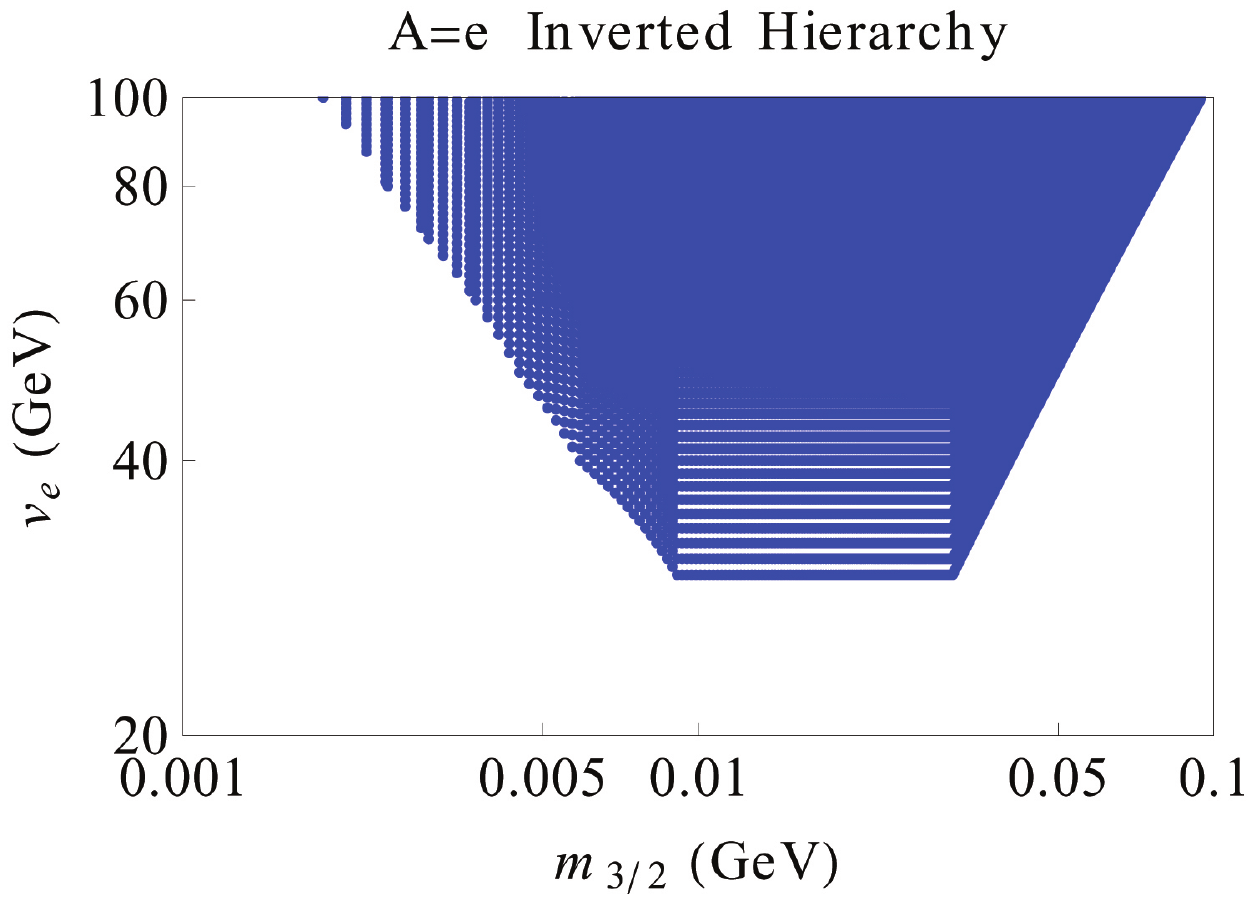}  \\
  \end{tabular}
\caption{\label{fig:el-Higgs} Allowed region (colored) in parameter space for the flavor assignment $A=e$, $B=\mu$ and $C=\tau$ in the case of Inverted Hierarchy.}
 \end{figure}
\end{center}

\subsection{Neutrino physics in AMRB}\label{sec:AMRB_neutrinos}

Inspecting Eq. (\ref{eq:AMRB_L}), it is clear that the gaugino Majorana masses contribute to the neutralino-neutrino mass matrix. This resembles what happens in RPV theories with 
bilinear terms \cite{Barbier:2004ez}, 
where one neutrino gets a non zero mass already at tree level through its mixing with gauginos.
\par
On the contrary, in this scenario all neutrinos remain 
massless at tree level.
This is a striking difference with respect to the RPV case, and can be understood considering the approximate eigenstates of the neutrino mass matrix (calculated \emph{e.g.} 
using the usual seesaw formula):
\be\label{eq:neutr-eig-AM}
\begin{pmatrix}
\nu_A' \cr \nu_B'\cr \nu_B'
\end{pmatrix}
\simeq 
\begin{pmatrix}
\nu_A+ \frac{g v_A v_u \lambda_T}{\sqrt{2} \mu M_{\tilde W}}\tilde{h}_u +\frac{g' v_A}{\sqrt{2} M_B}\tilde{\psi}_B - 
\frac{g v_A}{\sqrt{2} M_{\tilde W}}\tilde{\psi}_W  \cr \nu_B\cr \nu_C
\end{pmatrix}
\ee
The $B$ and $C$ flavors are by themselves approximate eigenstates and cannot get mass through a mixing with gauginos. At the same time, the flavor $A$ mixes only with 
Higgsinos and adjoint fermions, so that the absence of mixing with gauginos and of Majorana masses for the adjoint fermions prevents $\nu_A$ from getting a tree level mass.
\par
It is now clear that, in the AMRB scenario, the only possibility for neutrinos to acquire a mass is through loop effects.
In the $(\nu_A, \nu_B, \nu_C)$ basis, the main contributions at 1-loop are given by \cite{Barbier:2004ez}:
\begin{itemize}
 \item Loops with two supersymmetric trilinear couplings and one mass insertion in the scalar propagator due to Anomaly Mediation.\\
Since this term is proportional to the mass of the fermion circulating in the loop, the dominant contributions are given by bottom quark, strange quark and tau lepton \footnote{We neglect 
the muon contribution  because, due to color factors, it is subdominant with respect to the strange quark contribution.}.
\par
They are given by:
\bea\label{eq:nu_masses_bottom}
m_\nu^{q} &=& \frac{6}{(16 \pi^2)^2}\left( \frac{m_{3/2} \ v_A}{m_{\tilde b}^2}\right) \hat{\beta_b} \left[
m_b \begin{pmatrix}
\lambda^{'2}_{A33} & \lambda'_{A33} \lambda'_{B33} &\lambda'_{A133} \lambda'_{B33} \cr
\lambda'_{133} \lambda'_{B33} & \lambda^{'2}_{B33} &\lambda'_{C33} \lambda'_{B33} \cr
\lambda'_{A33} \lambda'_{C33}&\lambda'_{C33} \lambda'_{B33} & \lambda^{'2}_{C33} \cr
\end{pmatrix}+ \right. \nonumber\\
&& \left. 
m_s \begin{pmatrix}
0 & 0 & 0 \cr
0 & \lambda^{'2}_{B32} & \lambda'_{B32} \lambda'_{C32} \cr
0 &\lambda'_{B32} \lambda'_{C32} & \lambda^{'2}_{C32} \cr
\end{pmatrix}\right]\;,
\eea
where $\lambda^{'}_{Aii}=(m_d)_i/v_A$ is the $i^{th}$ down-quark Yukawa coupling, $m_{\tilde{b}}$ is the common left handed and right handed sbottom mass scale, 
$\hat{\beta}_b$ is the bottom $\beta$-function \cite{Gherghetta:1999sw}, 
and for simplicity we have assumed $\lambda'_{B23}= \lambda'_{B32}$, $\lambda'_{C23}= \lambda'_{C32}$.
\par
In the lepton sector, the main contribution is given by
\be
m_\nu^{\tau}= \frac{2}{(16 \pi^2)^2}\left( \frac{m_{\tau} m_{3/2}\ v_A}{m_{\tilde \tau}^2}\right)  \hat \beta_{\tau}
\begin{pmatrix}
 \lambda_{A33}^2 &  \lambda_{A33}  \lambda_{B33} & 0 \cr
  \lambda_{A33}  \lambda_{B33}&  \lambda_{B33}^2  & -\lambda_{B33} \lambda_{C33} \cr
0 & -\lambda_{B33} \lambda_{C33} &  0 \cr
\end{pmatrix}\; ,
\ee
where $\lambda_{A33}= m_\tau/v_A$ is the tau Yukawa coupling and $\hat \beta_\tau$ the tau $\beta$-function \cite{Gherghetta:1999sw}.
 \item Loops with two gauge couplings and one Majorana mass insertion in the gaugino propagator:
\be\label{eq:nu_masses_gaugino}
(m^{\rm gg}_{\nu})_{AA}  = \frac{g^4}{4} \frac{m_{3/2}}{(16\pi^2)^2} \left(\frac{v_A}{v}\right)^2\frac{m_Z^2}{M_{\tilde W}^2}\;,
\ee
where $M_{\tilde W}$ is the Dirac Wino mass and we have used the Anomaly Mediation contribution to the Majorana Wino mass: $m_{\tilde W} = \frac{g^2}{16 \pi^2} m_{3/2}$.
\end{itemize}
In the previous equations we neglected the mixing of $\nu_a $ with the adjoint gauginos (see Eq. 
(\ref{eq:neutr-eig-AM})): this 
is consistent in the portion of parameter space we will consider in the following numerical analysis.
\par
Barring special relationship between the parameters involved, the neutrino mass matrix has now three non zero eigenvalues. These depend on free parameters 
(trilinear RPV couplings and gravitino mass), that can be chosen to fit the experimental data, but also on gauge couplings and masses which are constrained by collider experimental bounds
\footnote{In what follows we will always take as reference a ``natural'' spectrum for the supersymmetric partners, with only the squarks of the third generation below the TeV scale, 
while all other superparticle masses can be above the TeV scale. At the moment, the experimental bounds on this kind of spectrum is less severe than those obtained for 
almost degenerate squarks \cite{Papucci:2011wy}.
Note that Dirac gauginos have an improved naturalness with respect to Majorana gauginos \cite{Fox:2002bu}, and this allows us to have a natural gluino above the TeV scale and
a heavier Wino. In what follows, we will take the Dirac Wino mass up to $10\;\mathrm{TeV}$.}.
\par
As already stressed, our scenario is a particular case of RPV SUSY (in particular the loop contributions are the same in both cases), so that it is interesting to compare the two 
situations.
Usually in RPV scenarios the left/right sparticle mixing and the Majorana gauginos mass are at the EW scale, while in our case they are proportional to the gravitino 
mass and can be subleading for small supersymmetry breaking scale. This implies that while usually one needs to suppress too large loop contributions to neutrino masses putting severe upper bounds on the trilinear couplings \cite{Barbier:2004ez}, 
in our case the upper bound is translated on the gravitino mass (with trilinear couplings usually allowed to saturate the bounds from EWPM and flavor physics, see Sec. \ref{sec:exp-constr}).
\par
A loose upper bound on the gravitino mass can be derived from cosmological considerations.
 Indeed,  as already stressed, the absolute neutrino mass scale is bounded from above from CMB measurements, $m_\nu \lesssim 0.6\; \mathrm{eV}$. 
 This readily translates into an upper bound 
on the gravitino mass, which can be roughly estimate as follows. Since $m_{AA}$ is the only entry in the neutrino mass mass matrix that do not depend on trilinear couplings, we can 
use it to roughly set the largest neutrino eigenvalue scale. For typical value of sparticle masses ($m_{\tilde{b}, \tilde{\tau}}\lesssim 
1\;\mathrm{TeV}$, $M_{\tilde W} \lesssim 10\;\mathrm{TeV}$) we obtain
$$
m_{3/2} \lesssim 0.5\; \mathrm{GeV}.
$$
We will now study in detail  whether, in the AMRB scenario, the phenomenological neutrino mass matrices can be reproduced in the case where the flavor $A$ is the either electron, muon or tau. 

\begin{center}
 \begin{table}[tb]\centering
  \begin{tabular}{c|ccccc}
   $A=e$             & $|\lambda_{133}|$        & $v_A~(\mathrm{GeV})$ & $m_{\tilde{b}}~(\mathrm{GeV})$ & $m_{\tilde{\tau}}~(\mathrm{GeV})$ & $M_{\tilde W}~(\mathrm{TeV})$\\
  \hline
                     & $ 5\times 10^{-7} - 1.4$ &  $20 - 100$          & $300 - 1000$                   & $200 - 1000$                      & $0.5 - 10$ \\ 
                     &                          & & & \\ 
   $A=\mu$           & $|\lambda_{233}|$        & $v_A~(\mathrm{GeV})$ & $m_{\tilde{b}}~(\mathrm{GeV})$ & $m_{\tilde{\tau}}~(\mathrm{GeV})$ & $M_{\tilde W}~(\mathrm{TeV})$\\
  \hline
                     & $ 5\times 10^{-7} - 1.4$ &  $20 - 100$          & $300 - 1000$                   & $200 - 1000$                      & $0.5 - 10$ \\
		     &                          & 	   	       & 			                                            & \\ 
  $A=\tau$\emph{(i)} & 				& $v_A~(\mathrm{GeV})$ & $m_{\tilde{b}}~(\mathrm{GeV})$ & $m_{\tilde{\tau}}~(\mathrm{GeV})$ & \\
  \hline
                     &                          & $20 - 100$           & $300 - 1000$                   & $200 - 1000$                      &  \\
  $A=\tau$\emph{(ii)} &                         & $v_A~(\mathrm{GeV})$ & $m_{\tilde{b}}~(\mathrm{GeV})$ & $m_{\tilde{\tau}}~(\mathrm{GeV})$ & $M_{\tilde W}~(\mathrm{TeV})$\\
  \hline
                     &                          & $20 - 100$           & $300 - 1000$                   & $200 - 1000$                       & $0.5 - 10$ \\
\end{tabular}
\caption{\small \label{Tab:range} Range of parameters used in the scan of Secs. \ref{sec:electron}-\ref{sec:muon}.}
 \end{table}
\end{center}

\subsubsection{$A=e$: Electronic Higgs}\label{sec:electron}

In this case we assign $A=e,\; B=\mu,\; C=\tau$. We perform our numerical scan for the parameters of Table \ref{Tab:range}, requiring the other variables to 
reproduce the phenomenological matrices and imposing the constraints of \cite{Dreiner:2006gu}. For simplicity, we have assumed degeneracy between LH and RH sparticles, and a full family degeneracy in the 
slepton sector. Strictly speaking, this simplification implies that, barring accidental cancellations, a natural common slepton mass cannot be too large, since it enters in the determination of the Z 
mass through the 
minimization of the scalar potential. However, keeping in mind that only the LH slepton mass matrix affects the Higgs sector, and to have an idea of the general behavior of the model, 
we allow the common slepton mass to assume also larger values.
\par
The main result of this section is that while the Normal Hierarchy case can be reproduced only in a very small region of the parameter space (corresponding to 
$v_e \sim 100\;{\rm GeV}$ and rather large Dirac Wino masses, $M_{\tilde W} \gtrsim 5\;{\rm TeV}$), a much larger portion of parameter space is 
available for Inverted Hierarchy. This can be understood looking at the phenomenological matrices of Eq. (\ref{eq:exp_matr_NH}): the $m_{ee}$ entry in the Normal Hierarchy case 
is about one order of magnitude smaller than the one of the Inverted Hierarchy case. For this to happen, one needs 
large sneutrino vev and large Wino mass. This can be seen noting that we can parametrize $m_{ee}$ as
\begin{equation}\label{eq:mAA}
 m_{ee} \propto \left( \frac{\alpha}{v_e}+ \beta v_e^2\right) m_{3/2}
\end{equation}
where the first term comes from the squark and slepton loops ($\alpha \sim 1/ \tilde{m}^2$) while the second one is due to the Wino loop ($\beta \sim 1/M_{\tilde W}^2$). 
A large vev can suppress the first term, while a large Wino mass can suppress the largeness of the second one.
The available parameter space for Inverted Hierarchy is shown in Fig. \ref{fig:el-Higgs}, where the allowed region is the colored one. 
We do not show plots on the sparticle and Wino mass planes since these parameters are practically unconstrained. As can be seen, in the squark sector 
the diagonal trilinear couplings $\lambda'_{333}$, $\lambda'_{233}$ are rather small, both at most of order 
${\cal O}(10^{-2})$, while the off-diagonal trilinear couplings $\lambda'_{332}$, $\lambda'_{232}$ can be large, up to ${\cal O}(10^{-1})$. In the lepton sector 
we have again couplings $\lambda_{233},\; \lambda_{231}$ at most of order ${\cal O}(10^{-1})$.
\par
Another interesting consequence of  our analysis is that we can  set a more precise range on the gravitino mass,
\be\label{eq:gr-vev-bounds}
1\;\mathrm{MeV} \lesssim m_{3/2} \lesssim 100\;\mathrm{MeV}, 
\ee
Furthermore, we also have an indication on the sneutrino vev: we can fit the neutrino mass matrix in our framework only if the sneutrino vev is somewhat large, $v_e \gtrsim 30\;{\rm GeV}$, 
\emph{i.e.} $\tan\beta \equiv \frac{v_u}{v_e} \lesssim 6$.
 Let us  also notice that for larger sneutrino vev, a larger gravitino mass is allowed. This can be understood from Eq. (\ref{eq:mAA}), 
from which it is clear that for small sneutrino vevs the term between brackets can be large, so that in general a small gravitino mass is needed to suppress this entry. On 
the contrary, for larger values of the vev the term between brackets is more suppressed, and a larger gravitino mass is allowed.
\par
A comment on the situation for larger lightest $m_{lightest}$ is in order. We have explicitly checked the situation for $m_{lightest} \simeq 0.1\;{\rm eV}$, finding that only in very a small region 
of parameter space the phenomenological neutrino mass matrix can be reproduced. However, let us stress that in this case approximately the same region of parameter space can reproduce both 
Hierarchies, since now the typical form of the mass matrix in the two cases is similar (Eqs. (\ref{eq:phen_NH}),(\ref{eq:phen_IH})).

\begin{center}
 \begin{figure}[htb]
  \begin{tabular}{cc}
   \includegraphics[width=0.45\textwidth]{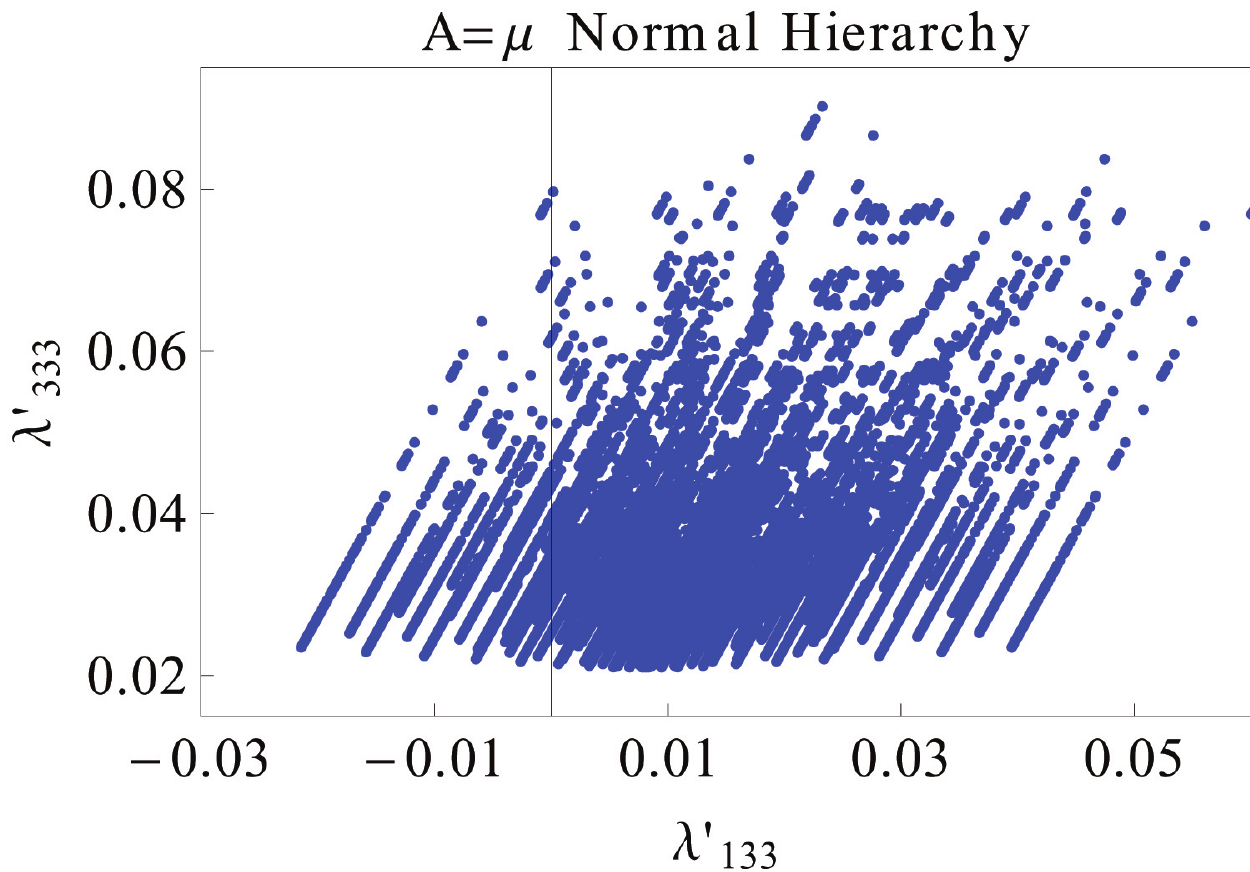} & \includegraphics[width=0.45\textwidth]{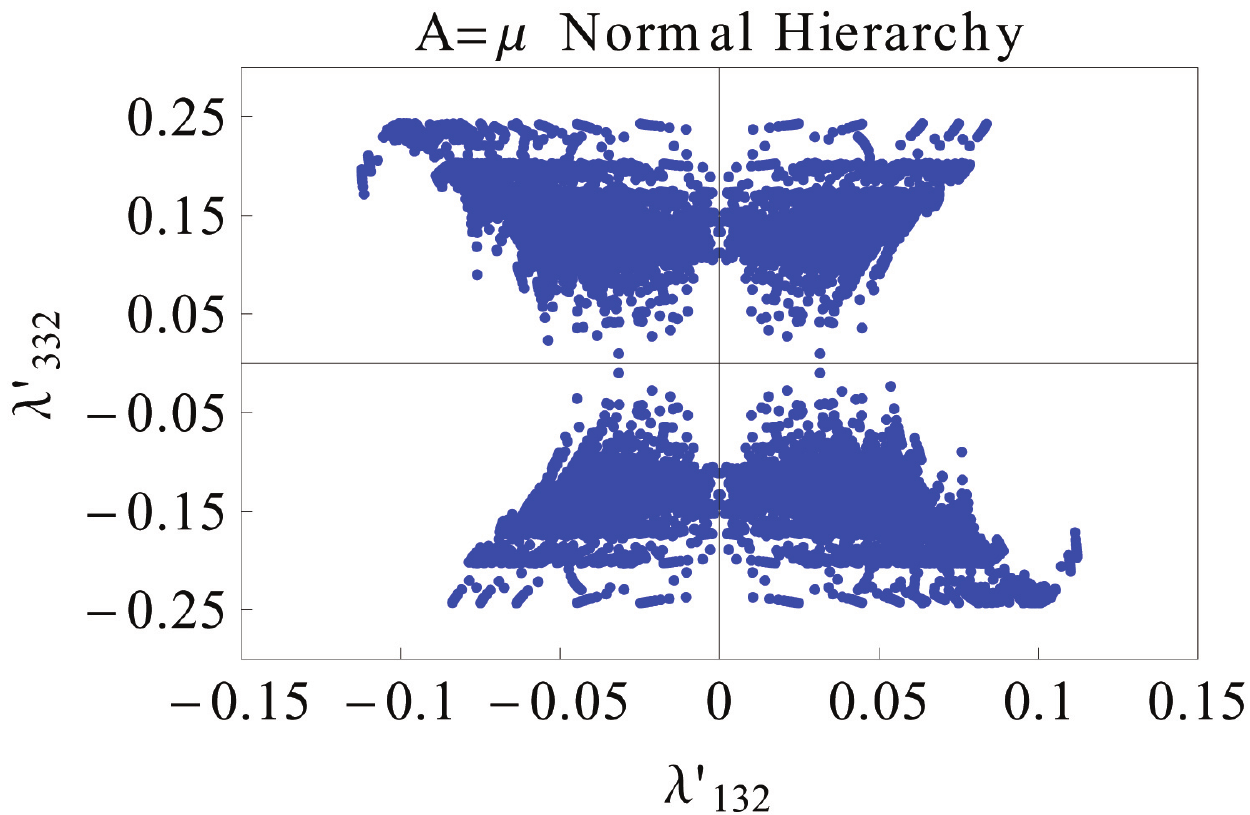} \\
   \includegraphics[width=0.45\textwidth]{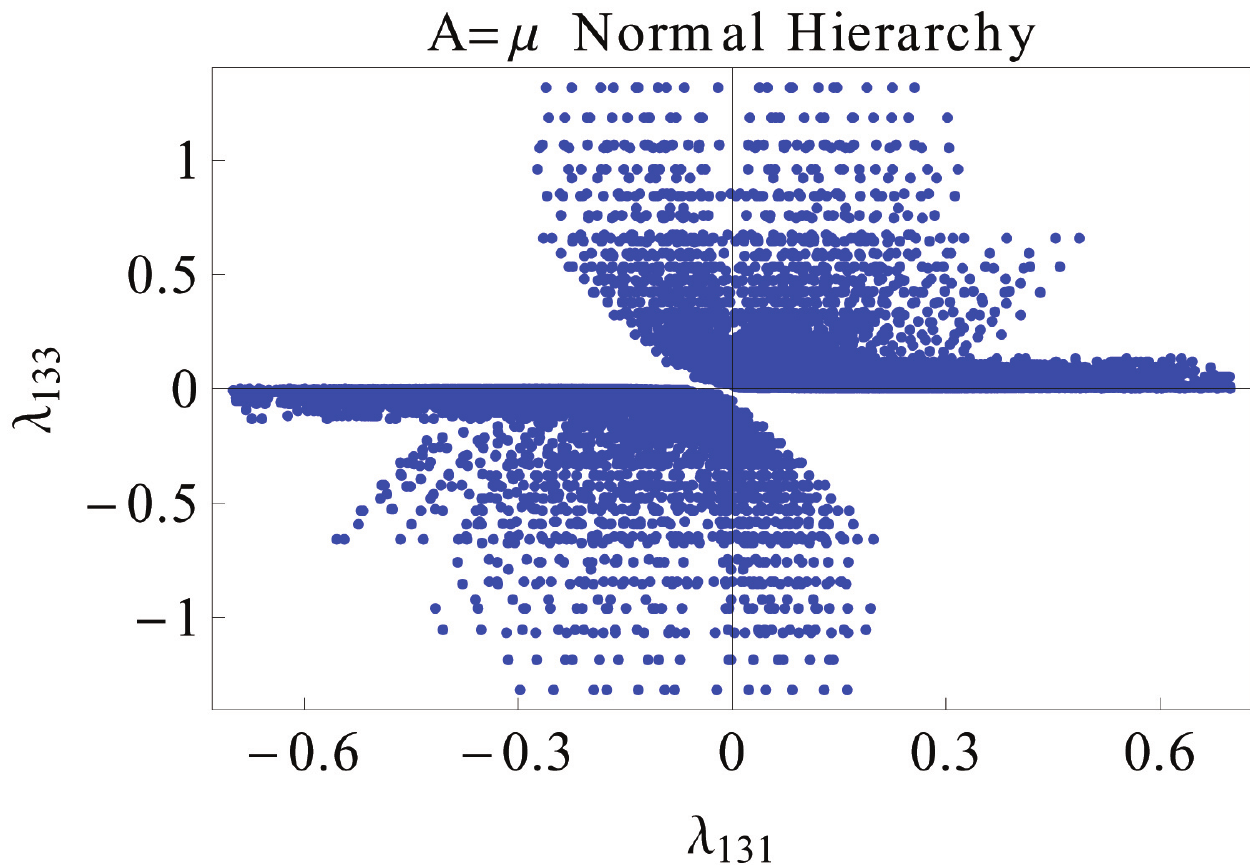} & \includegraphics[width=0.45\textwidth]{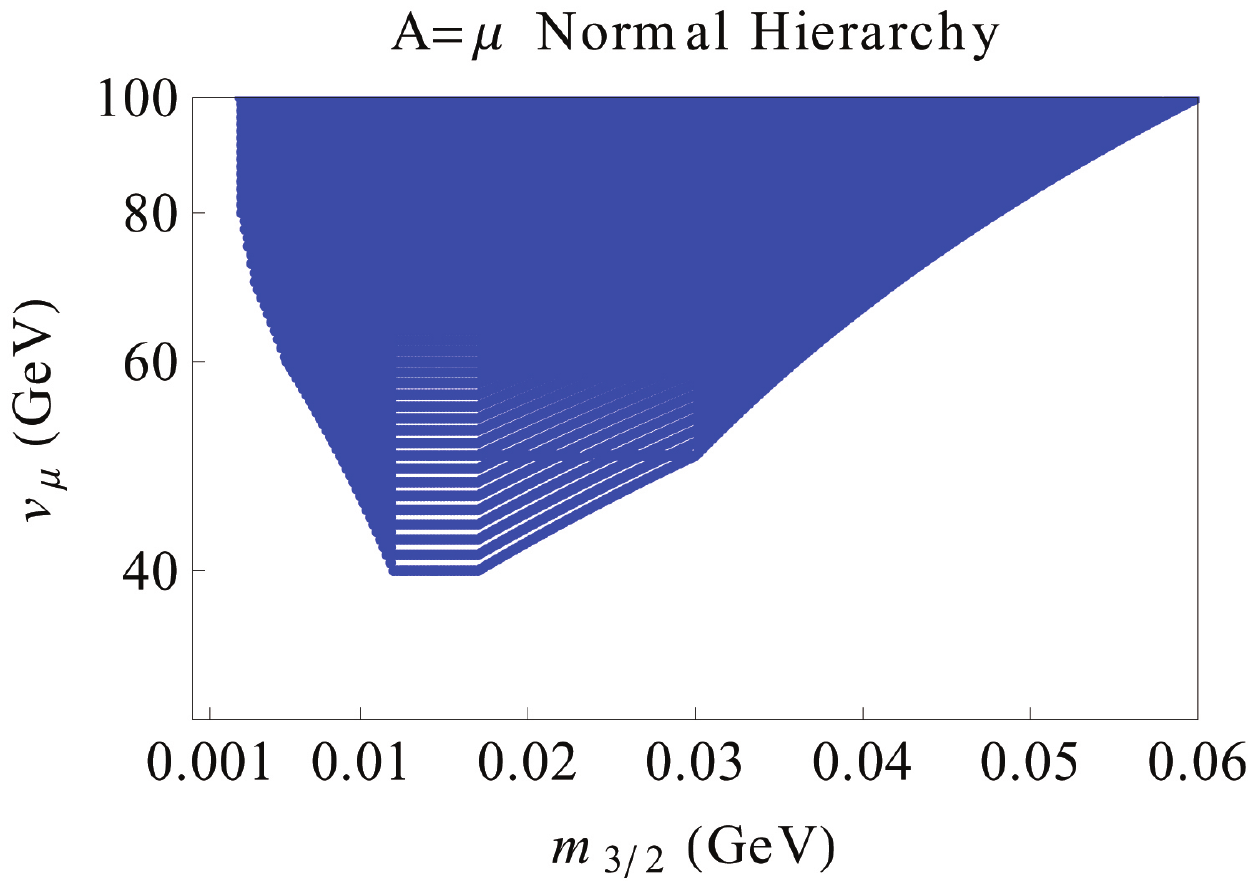} \\
  \end{tabular}
\caption{\label{fig:mu-Higgs} Allowed region (colored) in parameter space for the flavor assignment $A=\mu$, $B=e$ and $C=\tau$ in the case of Normal Hierarchy.}
 \end{figure}
\end{center}

\begin{center}
 \begin{figure}[htb]
  \begin{tabular}{cc}
   \includegraphics[width=0.45\textwidth]{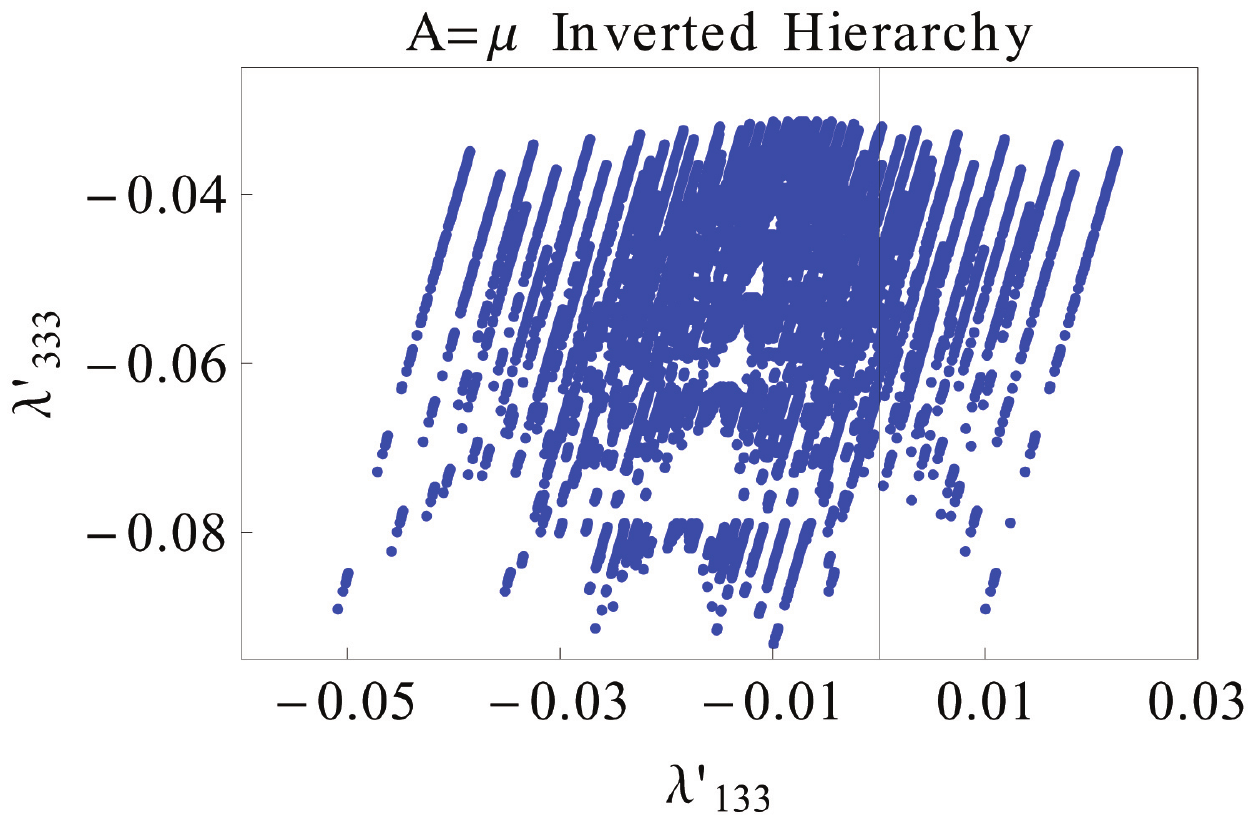} & \includegraphics[width=0.44\textwidth]{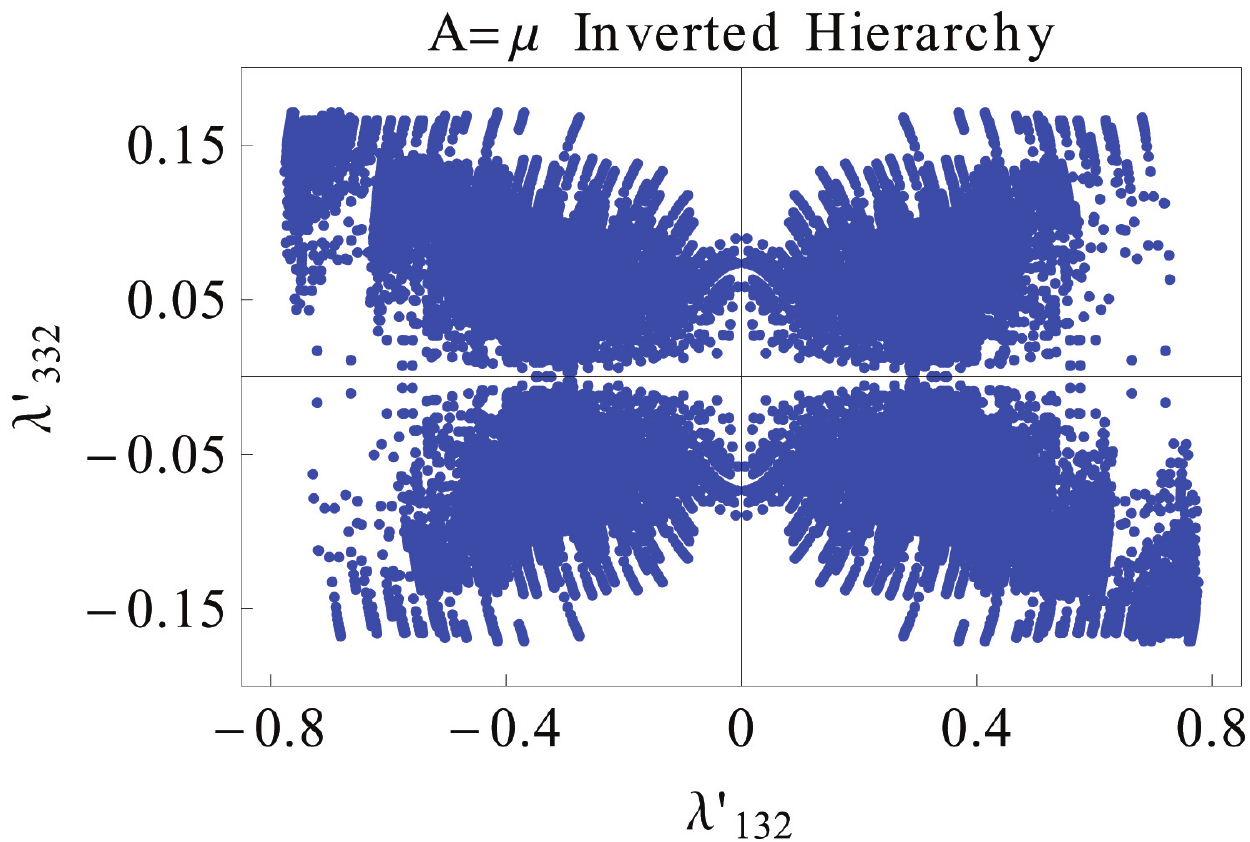} \\
   \includegraphics[width=0.45\textwidth]{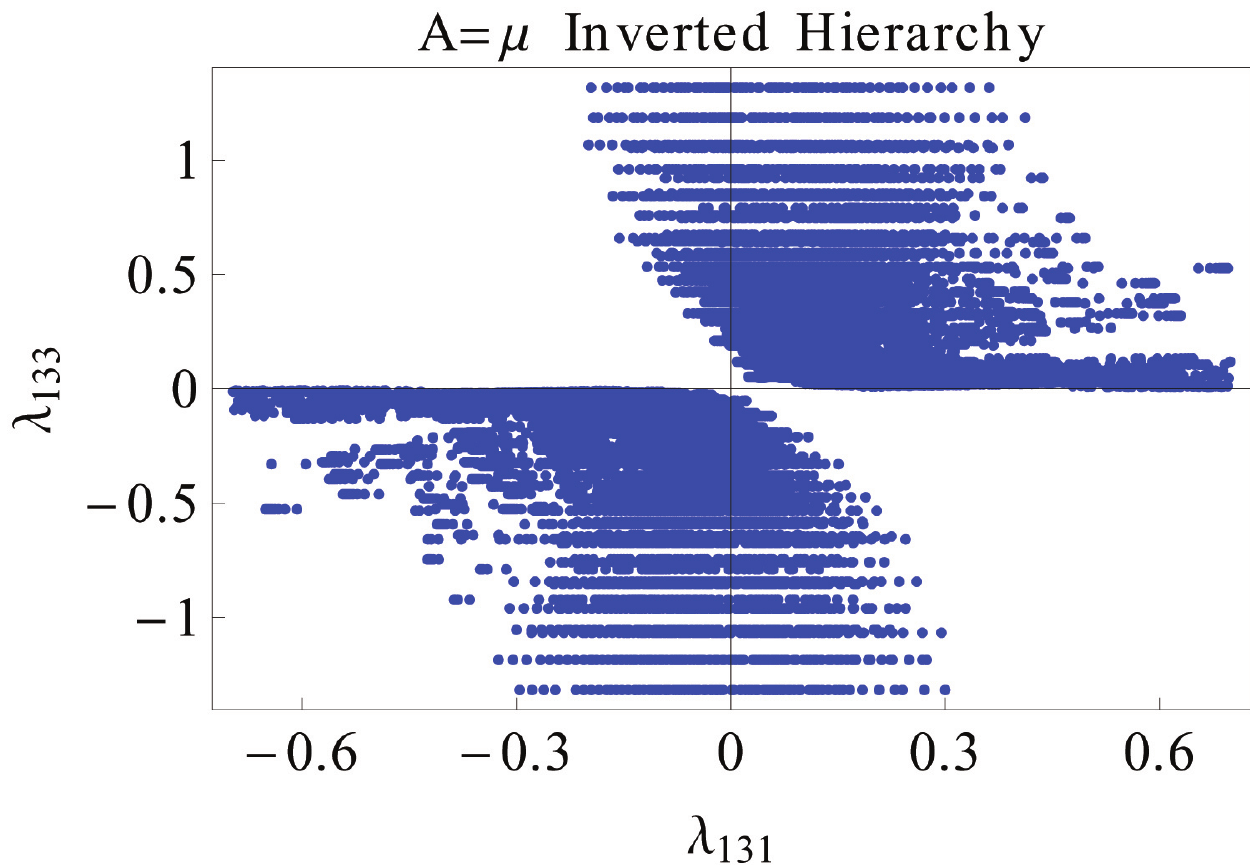} & \includegraphics[width=0.45\textwidth]{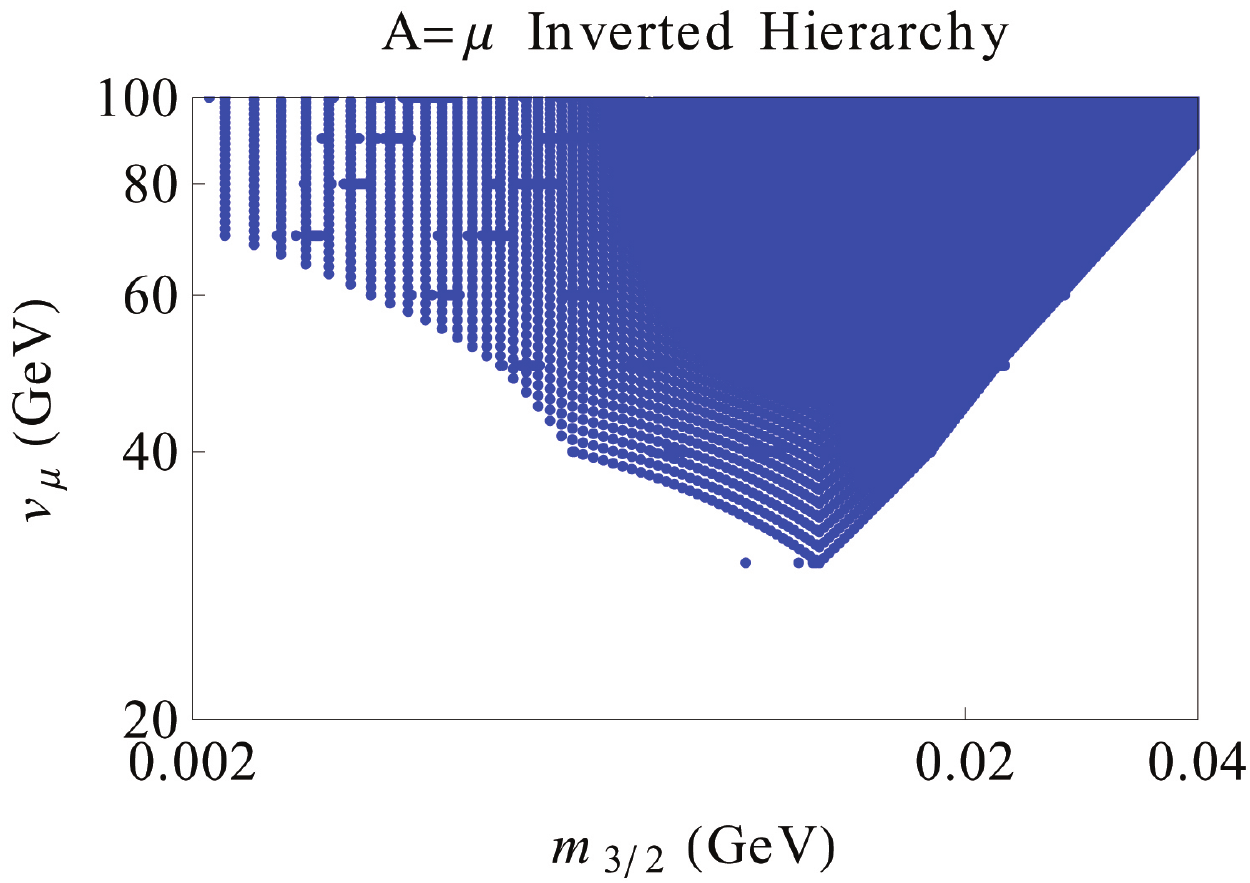} \\
  \end{tabular}
\caption{\label{fig:mu-Higgs_IH} Allowed region (colored) in parameter space for the flavor assignment $A=\mu$, $B=e$ and $C=\tau$ in the case of Inverted Hierarchy.}
 \end{figure}
\end{center}

\subsubsection{$A=\mu, \tau$: Muon and Tau Higgs}\label{sec:muon}

As pointed out in Sec. \ref{Sec:R-symm_as_Lepton_numb}, we consider the case of an Electronic-Higgs ($A=e$) more motivated from the point of view of the generation of the hard Yukawa couplings. 
However, for completeness 
we study also other possibilities. In particular, as we will see, the $A=\mu$ case offers an interesting different phenomenological situation with respect to the 
$A=e$ case.
\par
Let us start with $A=\mu$, $B=e$ and $C=\tau$. In this case Eq. (\ref{eq:mAA}) is valid for $m_{\mu\mu}$, which is similar for the two hierarchies. Thus, in general we expect that, unlike 
what happens in the $A=e$ case, both Hierarchies should be reproduced. This is indeed what happens, as confirmed by the scan performed for the parameters of 
Table \ref{Tab:range} within the same approximations described in the previous 
section. 
\par
The results are shown in Figs. \ref{fig:mu-Higgs}-\ref{fig:mu-Higgs_IH}. Also in this case we have checked that increasing the lightest neutrino mass diminishes drastically the available 
parameter space (although also in this case both Hierarchies can be accommodated).
\par
As can be seen from the plots, the range of parameters is roughly the same as the $A=e$ case, although some details can change. An exception is given by the lepton trilinear coupling 
$\lambda_{133}$, which is 
now allowed to be also of ${\cal O}(1)$. Regarding the muon-sneutrino vev and the gravitino mass, interestingly the situation does not change much with respect to the $A=e$ case: we conclude 
that the bounds of Eq. (\ref{eq:gr-vev-bounds}) are rather typical, for small neutrino masses, while they are no longer valid increasing the lightest neutrino mass.
\par
Let us now comment on the Tau-Higgs case, \emph{i.e.} $A=\tau$. We have performed our analysis both in the approximation of vanishing \emph{(i)} and non vanishing \emph{(ii)} muon mass.
In the case \emph{(i)} there is no contribution from loops involving sleptons, so that one can solve for the Dirac Wino mass instead of scanning on it. The results of our scan 
show that a solution compatible with the phenomenological mass matrices, Eq. (\ref{eq:exp_matr_NH}) requires either very large trilinear couplings or very large Wino masses (well above $100\;{\rm TeV}$). 
While the first possibility is excluded by the bounds coming from EWPM \cite{Dreiner:2006gu}, the second one is in principle viable. However, since as we already 
pointed out we want to stick to a spectrum 
which is not too unnatural, we consider this possibility at best marginal.\\
In the case \emph{(ii)} there is a non vanishing slepton loop contribution, in such a way that the scan on parameter space is quite similar to those of the two previous Sections (with the exception 
that in this case one of the two trilinear coupling constants involved is the muon-Yukawa coupling, so that there is no need to scan over it). Nevertheless, 
also in this case compatibility with Eq. (\ref{eq:exp_matr_NH}) requires trilinear couplings incompatible with the bounds of \cite{Dreiner:2006gu}.\\
The situation is summarized in Table \ref{Tab:range}; and the conclusion is that the case $A=\tau$ cannot reproduce neither a Normal nor a Inverted Hierarchy spectrum.

\subsection{Neutrino physics in PMRB}\label{sec:GMRB_neutrinos}

Let us now turn to the case where gravitational effects also break the ${\rm U(1)_R}$ symmetry \footnote{We thank T. Gr\'egoire, P. Kumar and E. Pont\'on for discussions on this whole section.}. 
The main difference with the previous case is that now two non zero neutrino masses are generated at tree level.
To understand this, let us consider the mixing among fermions in the neutralino sector.
In the R-symmetric limit, the $R=-1$ mass eigenstates are well approximated by:
\bea
 \nu'_A &\simeq &   \nu_A- \frac{ g v_A}{\sqrt{2} M_{W}} \psi_{\tilde W}+ \frac{ g' v_A}{\sqrt{2} M_{B}} \psi_{\tilde B}+ \frac{g v_A v_u \lambda_T}{\sqrt{2} \mu M_{\tilde W}}, \nonumber\\
 \psi_{{\tilde W}'}&\simeq & \psi_{\tilde W}-  \frac{ g v_A}{\sqrt{2} M_{W}}  \nu_A -  \frac{ g v_u}{\sqrt{2} M_B} \tilde h_u, \nonumber\\
 \psi_{{\tilde B}'}&\simeq & \psi_{\tilde B}-  \frac{ g' v_A}{\sqrt{2} M_{W}}  \nu_A -  \frac{ g v_u}{\sqrt{2} M_{B}} \tilde h_u,
 \eea
 while $\nu_B$ and $\nu_C$ do not mix, as we have already noticed. The  $R=1$ states are instead:
 \be
   \tilde W' \simeq \tilde W-  \frac{ \lambda_T}{\sqrt{2} M_{W}}  \tilde h_d, ~~~\tilde B' \simeq \tilde B-  \frac{ \lambda_S}{\sqrt{2} M_{B}}  \tilde h_d. 
 \ee
The inclusion of R-breaking effects generates new mixing terms for all neutrinos:
\bea
\nu'_A &\simeq & \nu_A- \frac{ g v_A}{\sqrt{2} M_{W}} \psi_{\tilde W}+ \frac{ g' v_A}{\sqrt{2} M_{B}} \psi_{\tilde B}+ \frac{g v_A v_u \lambda_T}{\sqrt{2} \mu M_{\tilde W}}- \frac{ \mu_A}{\mu} \tilde h_d , \nonumber \\
\nu_B &\simeq &  \nu_{B} - \frac{ \mu_B}{\mu} \tilde h_d, \nonumber\\
\nu_C &\simeq &  \nu_{B} - \frac{ \mu_C}{\mu} \tilde h_d,
\eea
which in turn produce a mass term for $\nu_A$ and mixing terms $m_{AB}$, $m_{AC}$:
\bea\label{eq:PMRB_treeA}
 m_{AA} &=& \sqrt{2} \frac{v_A v_u}{\mu} \left( \frac{g \lambda_T}{M_{\tilde W}}-\frac{g' \lambda_S}{M_B}\right) \mu_A , \nonumber \\
 m_{AB} &=& \frac{v_A v_u}{\sqrt{2} \mu}\left( \frac{g \lambda_T}{M_{\tilde W}} -\frac{g' \lambda_S}{M_B} \right) \mu_B , \nonumber\\
 m_{AC} &=& \frac{v_A v_u}{\sqrt{2} \mu}\left( \frac{g \lambda_T}{M_{\tilde W}} -\frac{g' \lambda_S}{M_B} \right) \mu_C.
\eea
Furthermore, a Majorana mass for the adjoint gauginos is  generated, and through it the neutrino $\nu_A$ acquire an additional mass term:
\bea
 m_A= ( \frac{ g v_A}{M_{W}} )^2 m_T+  ( \frac{ g' v_A}{M_{B}} )^2 m_S.
 \eea
 This is an example of inverse seesaw mechanism \cite{inverse_seesaw}, where the role of the right handed Dirac neutrinos is played by the Dirac gauginos.
Therefore, the tree level mass matrix in the PMRB scenario is:
\bea\label{eq:PMRB_tree}
 m_{AA} &=& \sqrt{2} \frac{v_A v_u}{\mu} \left( \frac{g \lambda_T}{M_{\tilde W}}-\frac{g' \lambda_S}{M_B}\right) \mu_A +\left(\frac{g' v_A}{M_B} \right)^2 m_S + \left(\frac{g v_A}{M_{\tilde W}}\right)^2 m_T , \nonumber \\
 m_{AB} &=& \frac{v_A v_u}{\sqrt{2} \mu}\left( \frac{g \lambda_T}{M_{\tilde W}} -\frac{g' \lambda_S}{M_B} \right) \mu_B , \nonumber\\
 m_{AC} &=& \frac{v_A v_u}{\sqrt{2} \mu}\left( \frac{g \lambda_T}{M_{\tilde W}} -\frac{g' \lambda_S}{M_B} \right) \mu_C,
\eea
which has indeed just a zero eigenvalue.
\par
Let us first of all discuss the upper bound on the gravitino mass imposed by the condition 
$m_{\nu} \lesssim 0.6\;{\rm eV}$.
Looking at the non zero entries of the mass  matrix we see that in general the upper bound depends on the value of $\lambda_{T,S}$. As in the AMRB case, we focus on the $m_{AA}$ entry. When 
the first term is negligible, the inverse seesaw term gives an upper bound $m_{3/2} \lesssim 1-10\;{\rm keV}$ for $M_{W}\simeq 1\;{\rm TeV}$ and $v_A \lesssim 100\;{\rm GeV}$. On the other 
hand, when the first term cannot be neglected, it dominates over the term coming from the inverse seesaw, and the upper bound now reads $m_{3/2} \lesssim \frac{0.1\;{\rm keV}}{\lambda_{S,T}}$ which 
can be more stringent than in the previous case (depending on the value of $\lambda_{T,S}$).
We conclude that, under these assumptions, in PMRB the upper bound on the gravitino mass can be significantly lower than the one of the AMRB scenario. 
\par
Let us now explain why in this case fitting neutrino physics calls for the introduction of a new sector in the model.
Inspecting the phenomenological mass matrices of Eq. \ref{eq:exp_matr_NH}, we see that both hierarchies require leading order entries in the $\mu-\tau$ sector, which cannot be 
accommodated by the mass matrix (\ref{eq:PMRB_tree}). This is true for any choice of the flavor $A$. At the same time, we expect loop factors to be much smaller than the 
tree level entries, so that the overall picture cannot be modified too much. This calls for the introduction of a new sector in the model.
We can wonder what is the minimal sector able generate neutrino masses and mixing.
First of all we would like to generate neutrino physics without the need for a new source of R-breaking.
This means we should consider a  mechanism that generates neutrino masses and mixing when the lepton number is broken at very low scale (the keV gravitino mass).
The minimal possibility we can think of is an inverse seesaw mechanism with additional electroweak singlets \footnote{In general, such Singlets may be present in the sector that generates 
the hard Yukawa coupling \cite{Frugiuele:2011mh}.}.
\newline
Therefore, we introduce a right handed Dirac neutrino (two singlets  $S$ and $ \bar S$ with $R=0$ and $R=2$ respectively) and the following terms in  the superpotential:
\be
W= \sum_i \lambda_i S H_u L_i+ M_S S  \bar S\;.
\ee
\par
Each singlet gets a Majorana mass of order of the gravitino mass trough R-breaking effects, and this generates a Majorana neutrinos mass of order  
$m_\nu \sim \frac{ \lambda_i  v_u}{M_S}  m_S $.
An interesting possibility  for the Dirac mass $M_S $ is the TeV scale, since this opens up a link between neutrino physics and LHC physics; however, 
a complete analysis of this situation is beyond the scope of the paper and we defer it to a future work.

\section{Conclusions}

With a luminosity of about $5\;{\rm fb^{-1}}$ already collected by the LHC, and without any hint of signal so far, the available parameter space of standard supersymmetric models is 
getting more and more constrained. 
This motivates the study of a larger portion of the weak scale supersymmetry landscape. Since neutrino physics can be a natural probe into new physics, it is natural to ask whether or 
not, given a specific framework, neutrino masses and mixing can be accommodated. In this work we have studied a supersymmetric scenario where a continuous 
R-symmetry is identified with the total Lepton Number, so that a possible connection to neutrino physics is immediate. In particular, we have found that neutrino physics is 
strongly connected with the mechanism of R-symmetry breaking, which in turn is related to supersymmetry breaking.
\par
When R-symmetry breaking effects are communicated to the visible sector solely via Anomaly Mediation, all neutrinos acquire  mass at  1-loop level. The hierarchy that can be 
reproduced depends crucially on the flavor of the sneutrino that gets a vev and plays the role of down type Higgs. For small values of the lightest neutrino mass, and for $A=e$, 
the case of Normal Hierarchy is disfavored, since 
it can be reproduced only in a very limited portion of the parameter space. On the contrary, for $A=\mu$, both hierarchies can be fitted in a consistent portion of parameter space. 
Finally, for $A=\tau$, we are not able to reproduce neutrino phenomenology solely via loop effects. The situation changes increasing the lightest neutrino mass, since in this case both 
hierarchies can be accommodated for $A=e$ and $A=\mu$ (but not for $A=\tau$), but only in a limited region of parameter space.
\par
Another possibility is that R-breaking effects are communicated to the visible sector at the Planck scale. In this case two non vanishing neutrino masses are generated at tree level, but 
with a pattern that does not allow to reproduce the phenomenological matrices studied. Since loop effects give subdominant contributions and cannot change the overall picture, we
conclude that a new sector must be added to the theory in order to reproduce neutrino physics. The minimal possibility is to introduce additional singlets (that can however be already 
present in the sector that generates the hard Yukawa couplings) in order to have an inverse seesaw 
mechanism. The study of this possibility is however beyond the scope of this paper.
\par
Since neutrino physics selects a particular region of the parameter space of the model, some consequences on Dark Matter and collider physics can be inferred. 
The cosmological upper bound on the sum of neutrino masses translates into an upper bound on  the gravitino mass, $m_{3/2} \lesssim 0.5\;{\rm GeV}$ for AMRB (with a more precise range selected 
by the neutrino mass matrix fit, $m_{3/2} \simeq 1\;{\rm MeV} - 100\;{\rm MeV}$)),  and $m_{3/2} \lesssim 10\;{\rm keV}$ for PMRB.
 In both scenarios the gravitino lifetime is long enough to evade all experimental bounds, so that it can be a Dark Matter candidate \cite{pheno}.
\par
Furthermore, neutrino physics selects also a preferred order of magnitude for the trilinear couplings both in the lepton and quark sector (with the general indication that 
the off diagonal couplings are larger that the diagonal ones). This can have important consequences for LHC physics. Indeed, one can expect squarks generation changing decays 
(as $\tilde{b}_L \rightarrow \nu_B s_R$ or $\tilde{t}_L \rightarrow e_B^+ s_R$) to dominate over the corresponding generation conserving decays 
($\tilde{b}_L \rightarrow \nu_B b_R$ or $\tilde{t}_L \rightarrow e_B^+ b_R$). 
A similar conclusion applies in the slepton sector, with decays like $\tilde{\nu}_B \rightarrow b\bar{s}$ or 
$\tilde{e}_B \rightarrow s\bar{t}$ generally dominating over $\tilde{\nu}_B \rightarrow b\bar{b}$ or $\tilde{e}_B \rightarrow b\bar{t}$. We defer to a future work \cite{pheno} the detailed 
analysis of possible signals.

\subsection*{Acknowledgements}
We thank Thomas Gr\'egoire, Piyush Kumar and Eduardo Pont\'on for valuable discussions and for carefully reading the manuscript. We thank Ben O'Leary for pointing out 
Ref. \cite{Dreiner:2006gu}.
E.B. acknowledges support from the Agence Nationale de la Recherche under contract ANR 2010 BLANC 0413 01.
C.F.  acknowledges support  by the Natural Sciences and Engineering Research Council of Canada
(NSERC).


\begin{thebibliography}{10}



\bibitem{Hall:1990dga}
L.~J. Hall, ``{ALTERNATIVE LOW-ENERGY SUPERSYMMETRY},''
\href{http://dx.doi.org/10.1142/S0217732390000536}{{\em Mod.Phys.Lett.}
  {\bfseries A5} (1990) 467}.

\bibitem{Hall:1990hq}
L.~Hall and L.~Randall, ``{U(1)-R symmetric supersymmetry},''
  \href{http://dx.doi.org/10.1016/0550-3213(91)90444-3}{{\em Nucl.Phys.}
  {\bfseries B352} (1991) 289--308}.

\bibitem{Barbier:2004ez}
R.~Barbier, C.~Berat, M.~Besancon, M.~Chemtob, A.~Deandrea, {\em et~al.},
  ``{R-parity violating supersymmetry},''
  \href{http://dx.doi.org/10.1016/j.physrep.2005.08.006}{{\em Phys.Rept.}
  {\bfseries 420} (2005) 1--202},
  \href{http://arxiv.org/abs/hep-ph/0406039}{{\ttfamily arXiv:hep-ph/0406039
  [hep-ph]}}.

\bibitem{Kribs:2007ac}
G.~D. Kribs, E.~Poppitz, and N.~Weiner, ``{Flavor in supersymmetry with an
  extended R-symmetry},''
  \href{http://dx.doi.org/10.1103/PhysRevD.78.055010}{{\em Phys.Rev.}
  {\bfseries D78} (2008) 055010},
  \href{http://arxiv.org/abs/0712.2039}{{\ttfamily arXiv:0712.2039 [hep-ph]}}.

\bibitem{Davies:2011mp}
  R.~Davies, J.~March-Russell and M.~McCullough,
  ``A Supersymmetric One Higgs Doublet Model,''
  JHEP {\bf 1104} (2011) 108
  [arXiv:1103.1647 [hep-ph]].

\bibitem{Frugiuele:2011mh}
  C.~Frugiuele and T.~Gregoire,
  ``Making the Sneutrino a Higgs with a $U(1)_R$ Lepton Number,''
  Phys.\ Rev.\ D {\bf 85} (2012) 015016
\href{http://arxiv.org/abs/1107.4634}{{\ttfamily  arXiv:1107.4634 [hep-ph]}}


\bibitem{Fox:2002bu}
P.~J. Fox, A.~E. Nelson, and N.~Weiner, ``{Dirac gaugino masses and supersoft
  supersymmetry breaking},'' {\em JHEP} {\bfseries 0208} (2002) 035,
  \href{http://arxiv.org/abs/hep-ph/0206096}{{\ttfamily arXiv:hep-ph/0206096
  [hep-ph]}}.

\bibitem{Papucci:2011wy}
M.~Papucci, J.~T. Ruderman, and A.~Weiler, ``{Natural SUSY Endures},''
  \href{http://arxiv.org/abs/1110.6926}{{\ttfamily arXiv:1110.6926 [hep-ph]}}.

\bibitem{Benakli:2011vb}
K.~Benakli, ``{Dirac Gauginos: A User Manual},''
  \href{http://arxiv.org/abs/1106.1649}{{\ttfamily arXiv:1106.1649 [hep-ph]}}.

\bibitem{Freitas:2009dp}
A.~Freitas, ``{Distinguishing Majorana and Dirac Gluinos and Neutralinos},''
  \href{http://dx.doi.org/10.1063/1.3327612}{{\em AIP Conf.Proc.} {\bfseries
  1200} (2010) 446--449}, \href{http://arxiv.org/abs/0909.5308}{{\ttfamily
  arXiv:0909.5308 [hep-ph]}}.

\bibitem{Heikinheimo:2011fk}
M.~Heikinheimo, M.~Kellerstein, and V.~Sanz, ``{How Many Supersymmetries?},''
  \href{http://arxiv.org/abs/1111.4322}{{\ttfamily arXiv:1111.4322 [hep-ph]}}.

\bibitem{Kribs:2012gx}
  G.~D.~Kribs and A.~Martin,
  ``Supersoft Supersymmetry is Super-Safe,''
  arXiv:1203.4821 [hep-ph].


\bibitem{Choi:2009jc}
S.~Choi, M.~Drees, J.~Kalinowski, J.~Kim, E.~Popenda, {\em et~al.},
  ``{Color-octet scalars at the LHC},'' {\em Acta Phys.Polon.} {\bfseries B40}
  (2009) 1947--1956, \href{http://arxiv.org/abs/0902.4706}{{\ttfamily
  arXiv:0902.4706 [hep-ph]}}.


\bibitem{Fok:2010vk}
R.~Fok and G.~D. Kribs, ``{mu to e in R-symmetric Supersymmetry},''
  \href{http://dx.doi.org/10.1103/PhysRevD.82.035010}{{\em Phys.Rev.}
  {\bfseries D82} (2010) 035010},
  \href{http://arxiv.org/abs/1004.0556}{{\ttfamily arXiv:1004.0556 [hep-ph]}}.


\bibitem{fayet}
  P.~Fayet,
  Nucl.\ Phys.\  {\bf B90 } (1975)  104-124.
  P.~Fayet,
  Phys.\ Lett.\  {\bf B64 } (1976)  159.
    P.~Fayet,
  Phys.\ Lett.\  {\bf B69 } (1977)  489.
  P.~Fayet,
  Phys.\ Lett.\  {\bf B78 } (1978)  417.


\bibitem{Gherghetta:2003wm}
  T.~Gherghetta and A.~Pomarol,
  Phys.\ Rev.\ D {\bf 67} (2003) 085018
  [hep-ph/0302001].



\bibitem{Brust:2011tb}
C.~Brust, A.~Katz, S.~Lawrence, and R.~Sundrum, ``{SUSY, the Third Generation
  and the LHC},''
\href{http://arxiv.org/abs/1110.6670}{{\ttfamily arXiv:1110.6670 [hep-ph]}}.

\bibitem{sneutrinovev}
C.~S.~Aulakh, R.~N.~Mohapatra,
  Phys.\ Lett.\  {\bf B119 } (1982)  136.
  L.~J.~Hall, M.~Suzuki,
  Nucl.\ Phys.\  {\bf B231 } (1984)  419.
  I-H.~Lee,
  Phys.\ Lett.\  {\bf B138}, 121 (1984).
    J.~R.~Ellis, G.~Gelmini, C.~Jarlskog, G.~G.~Ross, J.~W.~F.~Valle,
  Phys.\ Lett.\  {\bf B150 } (1985)  142.
  G.~G.~Ross, J.~W.~F.~Valle,
  Phys.\ Lett.\  {\bf B151 } (1985)  375.
  S.~Dawson,
  Nucl.\ Phys.\  {\bf B261 } (1985)  297.
 D.~E.~Brahm, L.~J.~Hall, S.~D.~H.~Hsu,
  Phys.\ Rev.\  {\bf D42 } (1990)  1860-1862.
   D.~-s.~Du, C.~Liu,
  Mod.\ Phys.\ Lett.\  {\bf A8 } (1993)  2271-2276.
     T.~Banks, Y.~Grossman, E.~Nardi, Y.~Nir,
  Phys.\ Rev.\  {\bf D52 } (1995)  5319-5325.
\href{http://arxiv.org/abs/hep-ph/9505248}{{\ttfamily hep-ph/9505248}}.


\bibitem{Kribs:2010md}
  G.~D.~Kribs, T.~Okui and T.~S.~Roy,
  ``Viable Gravity-Mediated Supersymmetry Breaking,''
  Phys.\ Rev.\ D {\bf 82} (2010) 115010
  [arXiv:1008.1798 [hep-ph]].

\bibitem{Gherghetta:1999sw}
T.~Gherghetta, G.~F. Giudice, and J.~D. Wells, ``{Phenomenological consequences
  of supersymmetry with anomaly induced masses},''
  \href{http://dx.doi.org/10.1016/S0550-3213(99)00429-0}{{\em Nucl.Phys.}
  {\bfseries B559} (1999) 27--47},
\href{http://arxiv.org/abs/hep-ph/9904378}{{\ttfamily arXiv:hep-ph/9904378 [hep-ph]}}.

\bibitem{Dreiner:2006gu}
  H.~K.~Dreiner, M.~Kramer and B.~O'Leary,
  ``Bounds on R-parity violating supersymmetric couplings from leptonic and semi-leptonic meson decays,''
  Phys.\ Rev.\ D {\bf 75} (2007) 114016
  [hep-ph/0612278].


\bibitem{Kumar:2009sf}
  A.~Kumar, D.~Tucker-Smith and N.~Weiner,
  ``Neutrino Mass, Sneutrino Dark Matter and Signals of Lepton Flavor Violation in the MRSSM,''
  JHEP {\bf 1009} (2010) 111
  [arXiv:0910.2475 [hep-ph]].


\bibitem{Davies:2011js}
R.~Davies and M.~McCullough, ``{Small neutrino masses due to R-symmetry
  breaking for a small cosmological constant},''
\href{http://arxiv.org/abs/1111.2361}{{\ttfamily arXiv:1111.2361 [hep-ph]}}.


\bibitem{Fogli:2011qn}
G.~Fogli, E.~Lisi, A.~Marrone, A.~Palazzo, and A.~Rotunno, ``{Evidence of
  $\theta_{13}\neq 0$ from global neutrino data analysis},''
  \href{http://dx.doi.org/10.1103/PhysRevD.84.053007}{{\em Phys.Rev.}
  {\bfseries D84} (2011) 053007},
  \href{http://arxiv.org/abs/1106.6028}{{\ttfamily arXiv:1106.6028 [hep-ph]}}.

\bibitem{Schwetz:2011qt}
T.~Schwetz, M.~Tortola, and J.~Valle, ``{Global neutrino data and recent
  reactor fluxes: status of three-flavour oscillation parameters},''
  \href{http://dx.doi.org/10.1088/1367-2630/13/6/063004}{{\em New J.Phys.}
  {\bfseries 13} (2011) 063004},
\href{http://arxiv.org/abs/1103.0734}{{\ttfamily arXiv:1103.0734 [hep-ph]}}.

\bibitem{Schwetz:2011zk}
T.~Schwetz, M.~Tortola, and J.~Valle, ``{Where we are on $\theta\_{13}$:
  addendum to 'Global neutrino data and recent reactor fluxes: status of
  three-flavour oscillation parameters'},''
  \href{http://dx.doi.org/10.1088/1367-2630/13/10/109401}{{\em New J.Phys.}
  {\bfseries 13} (2011) 109401},
\href{http://arxiv.org/abs/1108.1376}{{\ttfamily arXiv:1108.1376 [hep-ph]}}.

\bibitem{Abazajian:2011dt}
K.~Abazajian, E.~Calabrese, A.~Cooray, F.~{De Bernardis}, S.~Dodelson, {\em
  et~al.}, ``{Cosmological and Astrophysical Neutrino Mass Measurements},''
  \href{http://dx.doi.org/10.1016/j.astropartphys.2011.07.002}{{\em
  Astropart.Phys.} {\bfseries 35} (2011) 177--184},
\href{http://arxiv.org/abs/1103.5083}{{\ttfamily arXiv:1103.5083
  [astro-ph.CO]}}.

\bibitem{Bhattacharyya:2011zv}
G.~Bhattacharyya, H.~Pas, and D.~Pidt, ``{R-Parity violating flavor symmetries,
  recent neutrino data and absolute neutrino mass scale},''
  \href{http://dx.doi.org/10.1103/PhysRevD.84.113009}{{\em Phys.Rev.}
  {\bfseries D84} (2011) 113009},
\href{http://arxiv.org/abs/1109.6183}{{\ttfamily arXiv:1109.6183 [hep-ph]}}.

\bibitem{An:2012eh}
  F.~P.~An {\it et al.}  [DAYA-BAY Collaboration],
  ``Observation of electron-antineutrino disappearance at Daya Bay,''
  \href{http://arxiv.org/abs/1203.1669}{{\ttfamily arXiv:1203.1669 [hep-ex]}}.


\bibitem{inverse_seesaw}
D.~Wyler and L.~Wolfenstein,
Nucl.\ Phys.\  B {\bf 218}, 205 (1983).
R.~N.~Mohapatra and J.~W.~F.~Valle,
Phys.\ Rev.\  D {\bf 34}, 1642 (1986).
G.~C.~Branco, W.~Grimus and L.~Lavoura,
Nucl.\ Phys.\  B {\bf 312}, 492 (1989).
M.C.~Gonzalez-Garcia and J.W.F.~Valle, Phys.~Lett.~{\bf B216} (1989) 316.
J.~Kersten and A.~Y.~Smirnov,
  Phys.\ Rev.\  D {\bf 76}, 073005 (2007)
\href{http://arxiv.org/abs/0705.3221}{{\ttfamily arXiv:0705.3221 [hep-ph]}}.
A.~Abada, C.~Biggio, F.~Bonnet, M.~B.~Gavela and T.~Hambye,
  JHEP {\bf 0712} (2007) 061
\href{http://arxiv.org/abs/0707.4058}{{\ttfamily arXiv:0707.4058 [hep-ph]}}.
M.~Shaposhnikov,
Nucl.\ Phys.\  B {\bf 763}, 49 (2007)
\href{http://arxiv.org/abs/hep-ph/0605047}{{\ttfamily arXiv:hep-ph/0605047}}.
M.B.~Gavela, T.~Hambye, D.~Hernandez and P.~Hernandez,
JHEP {\bf 0909}, 038 (2009)
\href{http://arxiv.org/abs/0906.1461}{{\ttfamily arXiv:0906.1461 [hep-ph]}}.

\bibitem{pheno}
C. Frugiuele, T. Gr\'egoire, P. Kumar, E. Ponton, in preparation.

\end{thebibliography}
\end{document}